\documentclass[12pt, a4paper]{article}
\pdfoutput=1

\usepackage[height=25cm,width=16cm]{geometry}
\usepackage{amsfonts}
\usepackage{amsmath}
\usepackage{amssymb}
\usepackage{ascmac}
\usepackage{dcolumn}
\usepackage{bm,here}
\usepackage[sort&compress, numbers, merge]{natbib}
\usepackage{subfig}
\usepackage{comment}
\usepackage{ifpdf}
\usepackage{slashed}
\usepackage{colortbl}
\usepackage{xcolor}
 \usepackage[mathscr]{eucal}
\usepackage[utopia]{mathdesign}
\ifpdf        
  \usepackage{graphicx}     %   usepackage without driver option
  \usepackage[bookmarksopen,colorlinks=true,linkcolor=bblue,
  citecolor=ppink,urlcolor=ppink,linktocpage=false]{hyperref}
\else     % For (p)LaTeX + dvipdfmx
  \usepackage[dvipdfmx]{graphicx}     %   usepackage with driver option
  \usepackage[dvipdfmx,bookmarksopen,colorlinks=true,linkcolor=bblue,
  citecolor=ppink,urlcolor=ppink,linktocpage=false]{hyperref}
\fi

\usepackage{multicol}
\definecolor{red}{rgb}{1,0,0}
\definecolor{blue}{rgb}{0,0,1}
\definecolor{orange}{rgb}{1,0.5,0}
\definecolor{ppink}{rgb}{1,0.4,0.4}
\definecolor{bblue}{rgb}{0.284602,0.317763,0.963947}
\usepackage{framed}
\definecolor{shadecolor}{rgb}{0.95,0.95,0.95}
\newenvironment{claim}{\begin{shaded}\noindent\ignorespaces}{\end{shaded}}

\makeatletter
\@addtoreset{equation}{section}

\makeatother
\newcommand{\vev}[1]{ \left\langle {#1} \right\rangle }

\newcommand{\ket}[1]{ | {#1} \rangle }

\newcommand{\GEV}{\  {\rm GeV} }

\newcommand{\prn}[1]{\left( {#1} \right)}
\newcommand{\com}[1]{\left[ {#1} \right]}

\newcommand{\der}{\partial}  
\newcommand{\dd}{\mathrm{d}}
\newcommand{\Mpl}{M_{\rm Pl}}

\newcommand{\abs}[1]{\left\vert {#1} \right\vert}

\newcommand{\Max}{{\rm Max}}

\def\Mpl{M_{\rm pl}}

\begin{document}

%%%%%%%%%%%%%%%%%%%%%%%%%%%
%%%%%%%%%%% Title %%%%%%%%%%%
%%%%%%%%%%%%%%%%%%%%%%%%%%%
\begin{titlepage}

\begin{flushright}
UT 16-04\\
IPMU 16-0012
\end{flushright}

\vskip 3cm
\begin{center}
{%\Huge \bf 
\fontsize{30pt}{40pt}\selectfont \bfseries
Fate of Electroweak Vacuum\\[.7em]
during Preheating
}
%\vskip 1.2cm
\vskip .65in

{\large Yohei Ema$^{\lozenge}$,
Kyohei Mukaida$^{\blacklozenge}$,
Kazunori Nakayama$^{\lozenge,\blacklozenge}$
}

\vskip .35in
\begin{tabular}{ll}
$^{\lozenge}$&\!\! {\em Department of Physics, Faculty of Science, }\\
& {\em The University of Tokyo, Bunkyo-ku, Tokyo 133-0033, Japan}\\[.3em]
$^{\blacklozenge}$ &\!\! {\em Kavli IPMU (WPI), UTIAS,}\\
&{\em The University of Tokyo, Kashiwa, Chiba 277-8583, Japan}
\end{tabular}

%\vskip 1.5cm
\vskip .75in

\begin{abstract}  
\noindent
Our electroweak vacuum may be metastable 
in light of the current experimental data of the Higgs/top quark mass.
If this is really the case, high-scale inflation models require a stabilization mechanism of our vacuum during inflation.
A possible candidate is the Higgs-inflaton/-curvature coupling
because it induces an additional mass term to the Higgs during the slow roll regime.
However, after inflation, the additional mass term oscillates,
and it can destabilize our electroweak vacuum via production of large Higgs fluctuations
during the inflaton oscillation era.
In this paper, we study whether or not the Higgs-inflaton/-curvature coupling
can save our vacuum
by properly taking account of Higgs production during the preheating stage.
We put upper bounds on the Higgs-inflaton and -curvature couplings,
and discuss possible dynamics that might relax them.
\end{abstract}

\end{center}
\end{titlepage}

%%%%%%%%%%%%%%%%%%%%%%%%%%%
%%%%%%%%%%% TOC %%%%%%%%%%%
%%%%%%%%%%%%%%%%%%%%%%%%%%%
\tableofcontents
\thispagestyle{empty}
\newpage
\setcounter{page}{1}

%%%%%%%%%%%%%%%%%%%%%%%%%%%%%%%%
%%%%%%%%% Introduction and Summary %%%%%%%%%
%%%%%%%%%%%%%%%%%%%%%%%%%%%%%%%%
\section{Introduction and Summary}
\label{sec:intro_sum}

%%%%%%%%%%%%%%%%%%%%%%%%%%%%%%%%
%%%%%%%%%%%% Introduction %%%%%%%%%%%%
%%%%%%%%%%%%%%%%%%%%%%%%%%%%%%%%
\subsection{Introduction}
\label{sec:intro}

The current measurements of the Higgs and top quark masses
suggest that the Higgs quartic coupling flips its sign well below the Planck scale
if there is no new physics  
other than the Standard Model (SM)~\cite{Sher:1988mj,Arnold:1989cb,Anderson:1990aa,Arnold:1991cv,Espinosa:1995se,Isidori:2001bm,Espinosa:2007qp,Ellis:2009tp,Bezrukov:2009db,EliasMiro:2011aa, Bezrukov:2012sa,Degrassi:2012ry, Buttazzo:2013uya,Bednyakov:2015sca}.
It indicates that there may be a true vacuum at a large Higgs field value region, and 
our electroweak vacuum may not be absolutely stable.
Since its lifetime is much longer than the age of the Universe,
the electroweak vacuum is metastable for the best fit values of
SM parameters.\footnote{
	Higher dimensional operators may shorten 
	the lifetime of the electroweak 
	vacuum~\cite{Branchina:2013jra,Branchina:2014usa,Branchina:2014rva}.
}
In this paper, we take this situation seriously, 
assuming the metastable electroweak vacuum.

In the cosmological context, an interesting consequence of the Higgs metastability 
is that high-scale inflation has a tension 
with it~\cite{Espinosa:2007qp,Kobakhidze:2013tn,Fairbairn:2014zia,Hook:2014uia,Kamada:2014ufa,Herranen:2014cua,Kearney:2015vba,Espinosa:2015qea}.
This is because Higgs acquires fluctuations of the order of the Hubble parameter 
$H_{\text{inf}}$ during inflation.
As a result, the Higgs field falls into a lower potential energy region, and 
the electroweak vacuum decays if the inflation scale is high enough. 
Therefore, some stabilization mechanism of Higgs is necessary for high-scale inflation scenarios to
be consistent with the metastable electroweak vacuum.
 A leading candidate of such a stabilization mechanism
is the following Higgs-inflaton/-curvature interaction:
\begin{align}
	{\cal L}_\text{int} (\phi, h) 
	&= 
		\begin{cases}
			- \ \cfrac{1}{2}\ c^2 \phi^2 h^2, \\[1em]
			- \ \cfrac{1}{2}\ \xi R h^2,
		\end{cases}
		\label{eq:stb_intro}
\end{align}
where $\phi$ is the inflaton field, $h$ is a radial component of the Higgs field 
and $R$ is the Ricci scalar. 
Higgs acquires an effective mass term from this interaction during inflation, 
and hence the Higgs fluctuations are suppressed
if the conditions $c^2\phi_\text{inf}^2, \xi R_\text{inf} \gtrsim H_{\text{inf}}^2$ are satisfied.
Thus, the interaction~\eqref{eq:stb_intro} can stabilize the electroweak vacuum during inflation.

However, in such a case, the interaction~\eqref{eq:stb_intro} itself makes 
the dynamics during the preheating stage\footnote{
	A note on terminology.
	In this paper, the word ``preheating" represents 
	the epoch in which some resonant particle production processes occur
	due to the inflaton oscillation after inflation.
} highly nontrivial. 
The Higgs-inflaton quartic coupling causes 
the broad resonance of Higgs~\cite{Kofman:1994rk,Kofman:1997yn}
due to the breakdown of its adiabaticity and its Bose enhancement.
As a result, the fluctuations of Higgs grow exponentially.\footnote{
	See also Refs.~\cite{Dolgov:1989us, Traschen:1990sw, Shtanov:1994ce}.
} The Higgs-curvature coupling causes a tachyonic enhancement of Higgs~\cite{Bassett:1997az,Tsujikawa:1999jh,Herranen:2015ima}
because the curvature-induced effective mass squared becomes negative for some period
during one oscillation of the inflaton.
Thus, the exponential growth of Higgs fluctuations may force 
our electroweak vacuum to decay into the true one during the preheating stage.
The interaction~\eqref{eq:stb_intro} may eventually fail to save the electroweak vacuum.

In this paper, we focus on the dynamics of Higgs during the preheating
caused by the interaction~\eqref{eq:stb_intro}.
The main purpose of this paper is to investigate in what parameter space
the interaction~\eqref{eq:stb_intro} does not trigger 
the electroweak vacuum decay during the preheating stage.
We use both analytical and numerical methods to study the effects of
the broad/tachyonic resonance on the Higgs metastability.
We also consider the interactions between Higgs and radiation composed
of other SM particles during preheating.
In addition, we discuss possible dynamics of Higgs after the resonance shuts off, 
that is to say, after the preheating.
In the next sub-section, we summarize main results of this paper for the convenience of readers. 
We also give the organization of this paper there.

%%%%%%%%%%%%%%%%%%%%%%%%%%%%%%%%
%%%%%%%%%%%% Summary %%%%%%%%%%%%
%%%%%%%%%%%%%%%%%%%%%%%%%%%%%%%%
\subsection{Summary}
\label{sec:sum}
Here main results of this paper are summarized. 
We study evolution of Higgs during the preheating stage
driven by the interaction~\eqref{eq:stb_intro},
and obtain the parameter region in which Higgs remains in the electroweak vacuum during the preheating.
To be more specific, the electroweak vacuum survives the preheating stage 
if the couplings satisfy the following inequalities.
\begin{claim}
\textit{Upper bounds on Higgs-inflaton and -curvature couplings}:
\begin{align}
	c &\lesssim 10^{-4} \  \com{\frac{0.1}{\mu_{\text{qtc}}}}\ \com{\frac{m_\phi}{10^{13} \,\text{GeV}}},
	\label{eq:claim_qtc_sum1}
\end{align}
\begin{flushright}
\textit{for the quartic coupling case, and}
\end{flushright}
\begin{align}
	\xi &\lesssim 10 \ \com{\frac{2}{n_{\text{eff}}\,\mu_{\text{crv}}}}^2 \com{\frac{\sqrt{2}\,\Mpl}{\Phi_\text{ini}}}^2,
	\label{eq:claim_crv_sum}
\end{align}
\begin{flushright}
\textit{for the curvature coupling case.}
\end{flushright}
\end{claim}
\noindent
Here $m_{\phi}$ is the inflaton mass, $\Phi_{\text{ini}}$ is the initial inflaton amplitude at the beginning of its oscillation,
$\Mpl$ is the reduced Planck mass, $n_{\text{eff}}$ is an effective number of oscillation
[See the text below Eq.~\eqref{eq:number_crv}], $\mu_{\text{qtc}} \simeq 0.1$
and $\mu_{\text{crv}} \simeq 2$. See also Eqs.~\eqref{eq:claim_qtc} and \eqref{eq:claim_crv}.
For simplicity, we assume that inflaton oscillates with a quadratic potential to derive these bounds.
We obtain the upper bounds from the requirement that 
the Hubble expansion should kill the broad/tachyonic resonance
before Higgs escapes from the electroweak vacuum.
See Sec.~\ref{sec:higgs_inflaton} for details.
In order to confirm these upper bounds, we perform classical lattice simulations.
Effects of other SM particles are discussed in Sec.~\ref{sec:higgs_radiation} 
and also in the first part of Sec.~\ref{sec:reheating}.
The couplings $c$ and $\xi$ should satisfy
$c \gtrsim {\cal O}(H_\text{inf}/\Phi_{\text{ini}})$ and 
$\xi \gtrsim {\cal O}(0.1)$~\cite{Espinosa:2007qp,Espinosa:2015qea}
to suppress the Higgs fluctuations during inflation.
Thus, our result indicates that 
the allowed values of the couplings lie in a rather narrow band,
for the stability of Higgs during both the inflation and the preheating stages.

After the preheating, the dynamics of Higgs becomes quite complicated once
we include the interactions among Higgs, top quark and SM gauge bosons.
If we ignore this interaction, the bounds 
on the couplings become severer than Eqs.~\eqref{eq:claim_qtc_sum1} and \eqref{eq:claim_crv_sum}.
This is because the cosmic expansion reduces the Higgs effective mass induced from Eq.~\eqref{eq:stb_intro}
much faster than the tachyonic mass induced from its self interaction [See Eqs.~\eqref{eq:claim_qtc_aft} and \eqref{eq:claim_crv_aft}].
Thus, it is essential to include the interaction among Higgs and SM particles so as to discuss the fate of electroweak vacuum
after the preheating.
One possible fate is that, after the electroweak vacuum survives the preheating stage, 
Higgs thermalizes with the other SM particles.
Once it is thermalized, the life time of our vacuum can be estimated by means of
the bounce method under a periodic Euclidean time~\cite{Affleck:1980ac,Linde:1981zj}.
For thermalized radiation, the electroweak vacuum 
does not decay after the preheating even with a temperature $T \sim \Mpl$ for the central value
of the top quark mass~\cite{ATLAS:2014wva}.
See  discussion in Sec.~\ref{sec:turb}.
Also, the direct decay of inflaton into other SM particles,
which is responsible for the complete reheating, would play essential roles. 
Typically, a larger reheating temperature tends to stabilize the electroweak vacuum,
while a smaller one does not play the role.
See discussion in Sec.~\ref{sec:reheating}.
Also, resonant production of other SM particles during the early stage of complete reheating, if any,
might affect the bounds given in Eqs.~\eqref{eq:claim_qtc_sum1} and \eqref{eq:claim_crv_sum}.
However, above mechanisms strongly depends on thermalization processes.
A detailed study on the dynamics of Higgs
after the preheating, including Higgs-radiation and inflaton-radiation
couplings, is left for a future work.

The organization of this paper is as follows. 
In Sec.~\ref{sec:higgs_inflaton}, we first shut off the Higgs-radiation couplings
and focus on the Higgs-inflaton/-curvature coupling during 
the preheating regime. We study effects of the broad/tachyonic resonance
on the electroweak vacuum stability both analytically and numerically.
Then, in Sec.~\ref{sec:higgs_radiation}, we turn on the Higgs-radiation coupling,
and investigate how it could modify results of the previous sections.
We will see that it is expected to be less significant during the preheating.
In Sec.~\ref{sec:aft_prh}, we discuss a possible fate of the electroweak vacuum after the preheating.
At the end of Sec.~\ref{sec:aft_prh},
we discuss how the decay of inflaton into other SM particles could change the results.
The last section is devoted to the conclusion and the discussion.

%%%%%%%%%%%%%%%%%%%%%%%%%%%%%%%%
%%%%%%%%% Higgs-Inflaton Coupling %%%%%%%%%
%%%%%%%%%%%%%%%%%%%%%%%%%%%%%%%%
\section{Higgs-Inflaton Coupling}
\label{sec:higgs_inflaton}

The main goal of our study here is to estimate 
the Higgs-inflaton/-curvature coupling strength
below which the electroweak vacuum survives the preheating stage.
In particular, we will make it clear in what condition
the electroweak vacuum does not decay albeit the resonance occurs 
at the beginning of the inflaton oscillation.
In order to achieve this goal, we use both analytical and numerical methods. 
The essential point here is that the coupling cannot be arbitrarily small
in order to suppress the Higgs fluctuation during inflation.

First, let us explain our setup.
In this section, we shut off the Higgs-radiation couplings, 
{\it i.e.}\ gauge and top Yukawa couplings,
so as to clarify the role of Higgs-inflaton and -curvature couplings in the preheating stage.
We study the following model throughout this section:\footnote{
For simplicity, we treat Higgs as a one component field here. However, our constraints
derived from now are not sensitive to it because they depend only logarithmically on the number of components.
}
\begin{align}
	{\cal L} &= {\cal L}_\text{inf} (\phi) + {\cal L}_\text{Higgs} (h) + {\cal L}_\text{int} (\phi, h),
	\label{eq:lag}
\end{align}
where $\mathcal{L}_\text{inf}$, $\mathcal{L}_\text{Higgs}$ and $\mathcal{L}_\text{int}$
are the Lagrangian densities of the inflaton sector, the Higgs sector and the interaction between
inflaton/curvature and Higgs, respectively.
We take the inflaton sector as
\begin{align}
	{\cal L}_\text{inf} (\phi) &= 
		\frac{1}{2} \der_\mu \phi \der^\mu \phi - \frac{1}{2} m_\phi^2 \phi^2,
	\label{eq:L_inf}
\end{align}
where we have assumed that the inflaton potential around its origin is quadratic.
Note that, although the simple chaotic inflation model with a quadratic potential is excluded by observations~\cite{Ade:2015lrj},
it is possible to modify the large filed behavior to make it consistent with observations
(see \textit{e.g.}\ Refs.~\cite{Bezrukov:2007ep,Nakayama:2010kt,Destri:2007pv,Nakayama:2013jka,Kallosh:2013hoa,Kallosh:2013yoa}).
We implicitly assume this in the following.
The Higgs sector is given by
\begin{align}
	{\cal L}_\text{Higgs} (h) &= 
		\frac{1}{2} \der_\mu h \der^\mu h - \frac{1}{4} \lambda (\mu) h^4,
	\label{eq:L_higgs}
\end{align}
where $\mu$ is the renormalization scale.
Here we have dropped the negative Higgs mass squared 
since it is irrelevant for our following discussion.
We roughly approximate the Higgs self coupling $\lambda (\mu)$ as
\begin{align}
	\lambda (\mu) \simeq \tilde\lambda \text{sign}\prn{h_\text{max} - \mu}; 
	~~~\text{with}~~\tilde \lambda \simeq 0.01.
\end{align}
See Refs.~\cite{Buttazzo:2013uya,Bednyakov:2015sca}
for the running of the Higgs four point coupling.
The scale $h_\text{max}$, where the Higgs quartic coupling becomes negative,
significantly depends on the top quark mass.
For its current central value, it is given by $h_\text{max} \simeq 10^{10}$ GeV~\cite{Bednyakov:2015sca}.
For the interaction term which induces a sizable effective mass of Higgs during inflation,
we consider two different interactions as representative models:
\begin{align}
	{\cal L}_\text{int} (\phi, h) 
	&= 
		\begin{cases}
			- \ \cfrac{1}{2}\ c^2 \phi^2 h^2 &\cdots \text{quartic}, \\[1em]
			- \ \cfrac{1}{2}\ \xi R h^2 &\cdots \text{curvature}.
		\end{cases}
		\label{eq:stab}
\end{align}

Here we comment on the renormalization scale $\mu$.
There are two relevant physical scales in the following discussion,
\textit{i.e.} a typical momentum scale of preheating $p_\ast$ defined later
and the dispersion of Higgs $\sqrt{\langle h^2 \rangle}$\footnote{
	One might think of the Hubble parameter $H$ as another choice,
	for it would be the scale for the long wave length mode.
	Even if it is the correct choice, our results do not change
	as long as $H > h_\text{max}$ is satisfied for the first $\mathcal{O}(10)$ 
	times of inflaton-oscillation.
} 
since we deal with an inhomogeneous Higgs field owing to Higgs particle production.
Thus, it is nontrivial which one we should choose
as the renormalization scale 
for the vacuum decay triggered by the preheating dynamics.
Fortunately, however, our results are \textit{insensitive} to
this choice as long as $m_\phi \gg h_\text{max}$.
This is because, even if we set $\mu = \sqrt{\langle h^2\rangle}$
and hence start with positive $\lambda$, 
a sizable Higgs dispersion, $\langle h^2 \rangle \sim p_\ast^2 \gtrsim m_\phi^2$, 
is produced right after the first oscillation of inflaton
in the case of our interest.
Therefore, the self coupling $\lambda$
becomes negative just after the first oscillation.
As a result, the dynamics is the same as that with $\mu = p_\ast (\gtrsim m_\phi)$, or 
the case where $\lambda$ is negative from the beginning.
For a classical lattice simulation, 
we take the renormalization scale as $\mu = \Max\, [h, p_\ast]$,\footnote{
 	As can be inferred from Eqs.~\eqref{eq:ema_cond} and \eqref{eq:ema_cond_2},
	not only the long wave length mode but
	all the modes of Higgs below the typical scale $p_\ast$ are relevant
	when the vacuum decay is triggered. 
	(Precisely speaking, in the case of the curvature coupling, the scale is much larger; $q^{1/4} p_\ast$.)
	This is because the conditions [Eqs.~\eqref{eq:ema_cond} and \eqref{eq:ema_cond_2}]
	indicate that all the modes below $p_\ast$ overcome the pressure of spacial gradients of $ p_\ast^{-1}$
	owing to the tachyonic effective mass term.
	This is the reason why we take this criterion.
}
but readers should keep in mind the insensitivity of our result to this choice.\footnote{
	The rescatterings of produced Higgs particles via the ``positive'' four point coupling may
	kill the resonance, if $p_\ast^2 < \lambda \left< h^2 \right>$ under $\sqrt{\left< h^2 \right>} < h_\text{max}$.
	This is the case for $\lambda > p_\ast^2 / h_\text{max}^2 > m_\phi^2 / h_\text{max}^2$.
	In the second inequality, we have used $p_\ast > m_\phi$ in Eq.~\eqref{eq:np_cond} [or \eqref{eq:tac_cond}].
	It means that $h_\text{max}$ has to be much larger than $m_\phi$ at least
	for the perturbative $\lambda$.
	Since we are interested in the opposite case $h_\text{max} \ll m_\phi$, 
	we can safely just take $\lambda$ to be negative in the simulation.
}

In the following, we study the preheating epoch
driven by the Higgs-inflaton/-curvature interaction~\eqref{eq:stab} separately. 
We first discuss qualitative behavior of the system, and in particular 
clarify the condition where the Higgs field rolls down to the lower potential energy regime
than the electroweak vacuum.
Then, we show results of numerical simulations and confirm our qualitative understanding.
In the subsequent sections, we will turn on the Higgs-radiation coupling
and discuss how it affects the fate of the electroweak vacuum.
For that purpose, it is helpful to understand the dynamics qualitatively at first,
so that we can apply our understandings to more complicated systems.

%%%%%%%%%%%%%%%%%%%%%%%%%%%%%%%%
%%%%%%%%%%% Preliminaries %%%%%%%%%%%
%%%%%%%%%%%%%%%%%%%%%%%%%%%%%%%%
\subsection*{Preliminaries}
\label{sec:preliminaries}

Before moving to each case, we summarize common features of this system.
At first, we can safely neglect effects of Higgs fluctuations on the inflaton dynamics
since there are no particles right after the inflation.
Then, the inflaton obeys the following approximated solution:
\begin{align}
	\phi (t) \simeq \Phi (t) \cos \prn{ m_\phi t  }; ~~~ \Phi (t) = \frac{\Phi_\text{ini}} {a^{\frac{3}{2}}(t)},
	\label{eq:sol_inf}
\end{align}
where $a (t) \propto t^{2/3}$ is the scale factor,
and $\Phi (t)$ is an inflaton amplitude with $\Phi_\text{ini}$ being its initial value
at the onset of the inflaton oscillation.
Correspondingly, the dispersion relation of Higgs also oscillates with time,
and the Higgs field acquires fluctuations as we shall see below.

Let us start with a mode expansion of the Higgs field:
\begin{align}
	h(x) = \int \frac{\dd^3 k}{\com{2 \pi a(t)}^{3/2}} 
	\com{
		\hat a_{\bm{k}} h_{\bm{k}} (t) e^{i \bm{k} \cdot \bm{x}} + \text{H.c.}
	},
	\label{eq:mode_exp}
\end{align}
where $\bm{k}$ is a comoving momentum and $a (t)$ is the scale factor.
Neglecting interaction terms, we find the equation of motion for the wave function $h_{\bm k}$:
\begin{align}
	0 = \ddot h_{\bm {k}} (t) + \com{ \omega_{k; h}^2 (t) + \Delta (t)} h_{\bm k} (t),
	\label{eq:mode_eq}
\end{align}
where $\Delta \equiv - 9 H^2 /4 - 3 \dot H /2$,
$H \equiv \dot a / a$ is the Hubble parameter, and
$\omega_{k;h}(t)$ is the time dependent dispersion relation of Higgs.
Its time dependence originates from the Higgs-inflaton/-curvature coupling [Eq.~\eqref{eq:stab}].
The creation/annihilation operator $\hat a_{\bm k}$ satisfies
the following algebras:
$[ \hat a_{\bm k}, \hat a^\dag_{\bm k'} ] = \delta ( \bm{k} - \bm{k'} )$ and
$[ \hat a_{\bm k}, \hat a_{\bm k'} ] = [ \hat a^\dag_{\bm k}, \hat a^\dag_{\bm k'} ] = 0$.
We can always take the initial conditions such that
$h_{\bm k} (0) = 1/\sqrt{2 \omega_{k;h} (0)}$ and 
$\dot h_{\bm k} (0) = - i \sqrt{\omega_{k;h}(0) / 2}$
using the Bogolyubov transformation.
Then, the initial vacuum state is annihilated by the corresponding annihilation operator $\hat a_{\bm k}$ 
as $\hat a_{\bm k} \ket{0; \text{in}} = 0$.
Here we keep only the leading order WKB result with respect to the cosmic expansion.
See App.~\ref{app:mode_exp} for more details.

One can discuss particle production by means of Bogolyubov coefficients 
using the above mode expansion as done in literature.
Nevertheless, it is instructive to see the same physics in a different viewpoint, by only looking at correlators,
since the relation with outcomes of classical lattice simulations can be seen clearly.\footnote{
 	Also, it may be conceptually clear especially when the effects of inflaton/Higgs particle production becomes relevant,
	though it is equivalent after an appropriate reinterpretation.
}
We define two correlators
and their Fourier transforms~\cite{baym1962quantum,Chou:1984es,Calzetta:1986cq,Berges:2004yj,calzetta2008nonequilibrium}:
\begin{align}
	\vev{ \com{h(x),  h(y)}_+} &\equiv \int_{\bm{k}} \frac{e^{i \bm{k} \cdot \prn{ \bm{x} - \bm{y} }}}{\com{a (x_0) a (y_0 )}^{3/2} }  \
	G_{F; h} (x_0, y_0; \bm{k}),  \label{eq:stat}
	\\[.5em]
	\vev{ \com{h(x),  h(y)}_-} &\equiv \int_{\bm{k}} \frac{e^{i \bm{k} \cdot \prn{ \bm{x} - \bm{y}  }}}{\com{ a(x_0) a(y_0) }^{3/2}} \
	G_{\rho;h} (x_0, y_0; \bm{k}),
	\label{eq:spec}
\end{align}
where $\com{\bullet, \bullet}_\pm$ stands for the anti-commutator/commutator respectively,
expectation values, $\langle \bullet \rangle$, are taken by the initial vacuum state,
and we adopt the following shorthanded notation $\int_{\bm{k}} \equiv \int \dd^3k / (2 \pi)^3$.
These two propagators, $G_{F/\rho}$, are referred to as the statistical/spectral functions respectively.
The statistical function encodes the occupation number and
the spectral function reflects the spectrum of theory 
(\textit{i.e.}~location of poles for particle-like excitations and branch-cuts for continuum).
They are expressed in terms of wave functions if one neglects interaction terms:
\begin{align}
	G_{F/\rho ; h} (x_0, y_0; {\bm k}) = \com{  h_{\bm k} (x_0) h_{\bm k}^\ast (y_0) \pm h_{\bm k}^\ast (x_0) h_{\bm k} (y_0) }.
\end{align}
In classical lattice simulations, which we discuss later, 
one can compute the left hand side of Eqs.~\eqref{eq:stat} and \eqref{eq:spec}
by regarding the field variables as classical ones.
Note that there is no counterpart of 
the spectral function in classical lattice simulations,
and indeed the classical approximations are justified only for $|G_F| \gg |G_\rho|$~\cite{Aarts:2001yn}.
By using $G_F$, we can define the expectation value of 
energy density $ \vev{\hat T^{00}}$ as~\cite{Anderson:2005hi,Tranberg:2008ae}:
\begin{align}
	\vev{\hat T^{00}_h (x)}
	& = \int_{\bm{k}/a (x_0)}  \frac{1}{4} \com{ \der_{x_0}  \der_{y_0} + \omega_{k;h}^2 (x_0) } 
	\left.G_{F;h} (x_0 , y_0; \bm{k}) \right|_{y_0 \to x_0}
	- \prn{\text{Vac.}} 
	+\cdots \\[.5em]
	& = \int_{\bm{k}/a (x_0)}
	\frac{1}{2} \com{
		\abs{ \dot h_{\bm{k}} (x_0) }^2 + \omega^2_{k;h} (x_0) \abs{h_{\bm k} (x_0)}^2 - \omega_{k;h} (x_0)
	}
	+\cdots.
\end{align}
Here, again, we only keep the leading order term in the WKB approximation with respect to the cosmic expansion.
The term (Vac.) subtracts vacuum fluctuations.
This motivates the following definition of the comoving number density in the momentum space:
\begin{align}
	n_{k; h} (t) = \frac{1}{2 \omega_{k;h}(t)} \com{
		\abs{ \dot h_{\bm{k}} (t) }^2 + \omega^2_{k;h} (t) \abs{h_{\bm k} (t)}^2
	} - \frac{1}{2},
	\label{eq:number_comoving}
\end{align}
at the leading order in $H^2 / \omega_{k;h}^2$ expansion.
Note that the initial condition of the wave function $h_{\bm k}$
implies $n_{k;h}(0) = 0$.
The physical energy and number densities are given by 
$\int_{\bm{k}/a(t)} \omega_{k;h} n_{k ;h}$ and 
$\int_{\bm{k}/a(t)} n_{k ;h}$, respectively.

The break-down of adiabaticity plays the key role in the following discussion.
To see this, we take the time derivation of the number density of Higgs:
\begin{align}
	\dot{n}_{k;h} \sim \mathcal{O}\left(\dot{\omega}_{k;h}/\omega_{k;h}^2 \right)\omega_{k;h}n_{k;h}.
	\label{eq:boltzmannlike}
\end{align}
Here we have used the equation of motion [Eq.~\eqref{eq:mode_eq}],
and keep the leading order terms in $H^2 / \omega_{k;h}^2$ expansion.
Thus, the number density increases 
only if adiabaticity breaks down, 
\textit{i.e.}~$| \dot \omega_{k;h} / \omega_{k;h}^2 | \gtrsim 1$.
Indeed, the WKB solution of Eq.~\eqref{eq:mode_eq}
$
	h_{\bm k} (t) = e^ {- i \int^t d\tau \omega_{k;h} (\tau)} / \sqrt{2 \omega_{k;h} (t)}
$,
which is valid for the adiabatic region
$| \dot \omega_{k;h} / \omega_{k;h}^2 | \ll 1$,
leaves the number density unchanged from that of vacuum
up to corrections of ${\cal O} (| \dot \omega_{k;h} / \omega_{k;h}^2 | )$.\footnote{
 	The narrow resonance may be regarded as particle production proportional to this small number,
	$| \dot \omega_{k;h} / \omega_{k;h}^2 | \ll 1$.
}
Thus, the Higgs field acquires fluctuations for the non-adiabatic change of $\omega_{k;h } (t)$
induced by the Higgs-inflaton/-curvature coupling [Eq.~\eqref{eq:stab}].\footnote{
	At the onset of inflaton oscillation,
	Higgs fluctuations are also produced because the Hubble parameter changes
	non-adiabatically~\cite{Ford:1986sy} except for the conformal coupling $\xi = 1/6$.
	However, the inflaton oscillation provides more violent production of Higgs fluctuations
	as we will see.
}

%%%%%%%%%%%%%%%%%%%%%%%%%%%%%%%%
%%%%%%%%%% Quartic Stabilization %%%%%%%%%%
%%%%%%%%%%%%%%%%%%%%%%%%%%%%%%%%
\subsection{Preheating via quartic stabilization: $c^2 \phi^2 h^2$}
\label{sec:quart}

\subsubsection*{Qualitative discussion}
First let us consider the Higgs-inflaton quartic coupling case.
In this case, the dispersion relation of Higgs is given by
\begin{align}
	\omega_{k;h}^2 (t) = c^2 \Phi^2 (t) \cos^2 \prn{ m_\phi t }
	+ \frac{\bm{k}^2}{a^2(t)} 
	+ \delta m_{\text{self};h}^2 (t)
	\label{eq:disp_quart}
\end{align}
where $\delta m^2_{\text{self};h}$ represents a finite density correction to the Higgs mass term
from the Higgs self interaction, which we discuss later.
At first $\delta m^2_{\text{self};h}$ can be neglected since there are no Higgs particles right after the inflation.
Utilizing the conventional parameterization,
\begin{align}
	q (t) = \frac{c^2 \Phi^2(t)}{4 m_\phi^2}, ~~~
	A_k (t) = \frac{k^2}{m_\phi^2 a^2 (t)} + \frac{c^2 \Phi^2 (t)}{2 m_\phi^2},
\end{align}
we can reorganize the equation of motion [Eq.~\eqref{eq:mode_eq}] as follows
\begin{align}
	\com{\frac{\dd^2}{\dd (m_\phi t)^2} + A_k(t) - 2 q (t) \cos (2 m_\phi t) }h_{\bm{k}} (t) = 0.
	\label{eq:mathieu_eq}
\end{align}
Here again we keep leading terms in $H^2 / \omega_{k;h}^2$ expansion.
If one neglects the cosmic expansion, namely for constant $q$ and $A_k$,
it is nothing but the Mathieu equation.
Depending on parameters $q$ and $A_k$, 
the wave function grows exponentially or oscillates with a constant amplitude.

From now we focus on the effect of inflaton oscillation on the adiabaticity of Higgs.
Note that the cosmic expansion can be treated adiabatically,
{\it i.e.}~$m_\phi \gg H$\footnote{
	Strictly speaking, a small oscillating term remains in the scale factor $a$~\cite{Ema:2015dka}.
} after several oscillations of inflaton,
and hence we do not consider its effect on the adiabaticity of Higgs.
Let us quantify when
the inflaton oscillation becomes non-adiabatic for the Higgs by using the condition 
$| \dot{\omega}_{k;h}/\omega_{k;h}^2 | > 1$.
The left-hand-side tends to become large when the inflaton passes through $\phi \sim 0$
because the denominator gets smaller. Around $\phi \sim 0$, one can estimate
\begin{align}
	\abs{ \frac{\dot{\omega}_{k;h}}{\omega_{k;h}^2} }
	\sim \abs{\frac{c^2 \Phi^2 m_\phi^2 t}{(k^2 / a^2 + c^2 \Phi^2 m_\phi^2 t^2)^{3/2}}}
	\lesssim
	\frac{c \Phi m_\phi}{ k^2/a^2}.
\end{align}
The second inequality is saturated if one inserts $t \sim k / a / c \Phi m_\phi$.
If $m_\phi \gg c \Phi$ (\textit{i.e.}~$q \ll 1$), Higgs is produced within narrow bands $k /a \sim m_\phi (1 \pm q)$.
$m_\phi \gg c \Phi$ ensures the adiabaticity is not broken down.
Since Higgs is boson, induced emission effects of previously produced Higgs could
make the dynamics non-perturbative (\textit{i.e.}~narrow resonance~\cite{Dolgov:1989us, Traschen:1990sw, Shtanov:1994ce}), 
but usually the cosmic expansion soon kills such a resonant amplification.\footnote{
	The modes within the resonant band are efficiently red-shifted away if $q^2 m_\phi < H$.
	If this condition fulfills, the narrow resonance does not take place.
	Since we are mainly interested in the very early stage of preheating after chaotic inflation,
	we expect $H \sim m_\phi$. One can see that $q^2 m_\phi < H$ tends to be satisfied 
	for $q \ll 1$.
}
Hence, the produced amount of Higgs fluctuations during the narrow resonance is not so significant.
On the other hand, for $m_\phi < c \Phi$ (\textit{i.e.}~$q > 1$), the bands get broad,
and the adiabaticity is badly broken down for modes with $k^2 / a^2 \lesssim c \Phi m_\phi$.
This is the sign of non-perturbative Higgs production~\cite{Kofman:1994rk,Kofman:1997yn}.
To sum up, the Higgs fluctuations with $k / a \lesssim p_\ast$ are efficiently produced
if the following inequality holds:
\begin{align}
	p_\ast (t) \gtrsim m_\phi;~~~ p_\ast (t) \equiv \sqrt{ c m_\phi \Phi(t) },
	\label{eq:np_cond}
\end{align}
where $p_\ast (= k_\ast / a)$ is a characteristic physical momentum of Higgs particle production.

We assume that this inequality is satisfied at the onset of the inflaton oscillation
in the following discussion.
It indicates that the quartic coupling should satisfy $c\Phi_{\rm ini} > m_\phi$, or
\begin{align}
	c > 4 \times 10^{-6} \com{\frac{m_\phi}{1.5\times10^{13}\GEV}}\com{\frac{\sqrt{2}\,\Mpl}{\Phi_\text{ini}} },
	\label{eq:qtc_lowb}
\end{align}
where $\Phi_\text{ini}$ is the initial inflaton amplitude.
This inequality implies $c\Phi_{\rm ini} > H_{\rm inf}$, with $H_{\rm inf}$ being the Hubble scale at the end of inflation,
since $m_\phi \gtrsim H_{\rm inf}$ holds generically.
Thus the stabilization of the Higgs field during inflation is ensured under the assumption (\ref{eq:np_cond}).\footnote{
	If there is a hierarchy $m_\phi \gg H_{\rm inf}$ (or $\Phi_{\rm ini}\ll M_{\rm pl}$), it is possible to choose $H_{\rm inf} < c\Phi_{\rm ini} < m_\phi$.
	In such a case, the Higgs field is stabilized during inflation and also there is no violent Higgs production after inflation.
} 
Hereafter we consider this situation.

The Higgs mode whose momentum is below $p_\ast$ 
acquires fluctuations for each passage of $\phi \sim 0$,
and its mode function becomes
$
	h_{\bm k} (t) = [ \alpha_{k}(t) e^{ - i \int^t \omega_{k;h}} 
	+ \beta_{k} (t) e^{ i \int^t \omega_{k;h}} ] / \sqrt{2 \omega_{k;h} (t)}
$
outside the region in which the adiabaticity of Higgs is broken down.
Here $\alpha_k$ and $\beta_k$ are the Bogolyubov coefficients, satisfying $|\alpha_k|^2 - |\beta_k|^2 = 1$.
As explained below Eq.~\eqref{eq:boltzmannlike},
Higgs fluctuations are not produced for $| \alpha_k | = 1$ and $\beta_k = 0$.
A non-zero value of $\beta_k$ indicates the Higgs production.
After several times of the passages of $\phi \sim 0$, its number density grows exponentially due to
the Bose enhancement~\cite{Kofman:1994rk,Kofman:1997yn}:
\begin{align}
	n_h (t) 
	&=  \int_{\bm{k} / a(t)} n_{k ;h} (t) =  \int_{\bm{k} / a(t)} \abs{\beta_k (t)}^2
	\simeq  \int_{\bm{k} / a(t)} \frac{e^{2 \mu_k m_\phi t}}{2} \\[.5em]
	& \sim \frac{1}{32 \pi^2} \sqrt{ \frac{\pi}{2 \mu_\text{qtc} m_\phi t} }\ e^{2 \mu_\text{qtc} m_\phi t} p_\ast^3 (t),
\end{align}
where $\mu_k$ is a momentum dependent function,
and it has a maximum value $\mu_\text{qtc} \simeq \mathcal{O}(0.1)$ at $k\simeq k_\ast / 2$.
Here we have used the steepest descent method to evaluate the integral,
and estimated the second derivative of $\mu_k$ as $\mu''_{k_\ast} \sim 2 \mu_\text{qtc} / \delta k^2$
with $\delta k \sim k_\ast / 2$.

While the Higgs fluctuation continuously grows,
it induces effective mass corrections to inflaton and Higgs itself 
via the quartic interaction, $c^2 \phi^2 h^2$, and the self interaction, $\lambda h^4$,
respectively.
In our case, the self coupling $\lambda$ is larger than $c^2$, and hence
we just consider the latter effect.
The self coupling $\lambda$ is negative in the case of our interest
at least for $m_\phi > h_\text{max}$, which is fulfilled for the center value of the top quark mass
[See also the discussion below Eq.~\eqref{eq:stab}].
Therefore, the self coupling induces the tachyonic mass term:
\begin{align}
	\delta m^2_{\text{self};h} (t) 
	&= - 3 \tilde \lambda \int_{{\bm k}/a (t)} \frac{1}{2} G_{F;h} (t, t; {\bm k})
	= - 3 \tilde \lambda \int_{\bm{k} / a(t)} \abs{h_{\bm k} (t)}^2 \nonumber\\[.5em]
	&\simeq - 3 \tilde \lambda \int_{\bm{k} / a(t)} \frac{n_{k;h}(t)}{\omega_{k;h} (t)} \nonumber\\[.5em]
	& \sim  - 3 \tilde \lambda \frac{n_h (t)}{\omega_{k_\ast; h} (t)}.
\end{align}
This term forces the Higgs field to 
roll down to its true minimum.

Now let us derive the condition when Higgs rolls down to the true vacuum.
One might estimate it by simply comparing the effective masses
induced by the Higgs-inflaton coupling and the self coupling as
$\lvert \delta m_{\text{self};h}^2\rvert \gtrsim c^2\Phi^2$, 
\textit{but this is not correct.}
The important point here is that the effective mass induced by the Higgs-inflaton coupling
is oscillating, and hence Higgs infrared modes is tachyonic for some small time interval
at around when $\phi$ crosses
the origin even if $\lvert \delta m_{\text{self}:h}^2\rvert \ll c^2\Phi^2$.\footnote{
	In other words, we cannot take the oscillation average
	if the time scale of our interest, $\delta m_{\text{self};h}$, is faster than the oscillation period,
	$m_\phi^2 < \lvert \delta m_{\text{self};h}^2 \rvert$,
	which is the case we have to deal with in order to estimate the upper bound on $c$.
}
The time interval $\Delta t$ during which the inflaton-induced Higgs effective mass
is negligible compared to $\delta m_{\text{self};h}^2$ is estimated as
\begin{align}
	m_{\phi}^2q(m_{\phi}\Delta t)^2 \sim \lvert \delta m_{\text{self};h}^2 \rvert,
\end{align}
where $\phi \sim \Phi m_\phi \Delta t$ at around the potential origin.
%We estimate the growth rate during this time interval as
During this time interval, the tachyonic effective mass, $\delta m_{\text{self};h}$, 
can make the Higgs field grow.
The growth rate of the Higgs field during this time interval is estimated as
$|\delta m_{\text{self};h}| \Delta t \sim |\delta m_{\text{self};h}^2|/p_*^2$.
If it exceeds unity, the Higgs field increases by an large amount 
due to the tachyonic effective mass, $\delta m_{\text{self};h}$,
and it escapes from the metastable electroweak vacuum.
Thus, the following inequality should be satisfied
for the electroweak vacuum not to decay:
\begin{align}
	\abs{ \delta m_{\text{self};h}^2 (t) }_{\phi \sim 0} \lesssim p_\ast^2 (t)
	\leftrightarrow
	\frac{3 \tilde \lambda}{16 \pi^2} \sqrt{\frac{\pi}{2 \mu_\text{qtc} m_\phi t}} \ e^{2 \mu_\text{qtc} m_\phi t} 
	\lesssim 1,
	\label{eq:ema_cond}
\end{align}
where the subscript $\phi \sim 0$ indicates that the effective mass is evaluated
at the passage of $\phi \sim 0$.
Interestingly, the left inequality suggests that not only the long wave length mode of Higgs
but all the modes below $p_\ast$ overcomes the pressure due to the tachyonic effective mass
if this inequality is violated. Hence, the transition is expected to be dominated by the scale below $p_\ast$.

Eq.~\eqref{eq:ema_cond} implies that the electroweak vacuum decays at\footnote{
If $h_\text{max}$ is sufficiently large, say $h_\text{max} \gg (q/\tilde\lambda^2)^{1/4}m_\phi$, $t_{\rm dec}$ corresponds to the time 
at which the resonance shuts off due to the positive Higgs quartic coupling. 
Here we assume $h_{\rm max}$ is quite small: $h_{\rm max}\simeq 10^{10}\,$GeV.
}
\begin{align}
	t_\text{dec} \sim \frac{1}{2 \mu_\text{qtc} m_\phi} \ln \prn{ \frac{16 \pi^{\frac{3}{2}} }{3 \tilde \lambda} }.
	\label{eq:qrt_dec}
\end{align}
Before this time, the non-perturbative Higgs production should 
shut off due to the cosmic expansion for the electroweak vacuum to survive the preheating stage.
The cosmic expansion kills the non-perturbative Higgs production at $p_\ast (t_\text{end}) \sim m_\phi$,\footnote{
	Strictly speaking, after the broad resonance, a narrow resonance takes place,
	but it is soon killed by the cosmic expansion at $q^2 m_\phi \simeq H$.
}
which implies
\begin{align}
	t_\text{end} \sim \frac{2\sqrt{6}}{3 m_\phi} \ \frac{ c \Mpl}{m_\phi}.
	\label{eq:qrt_end}
\end{align}
Therefore, the electroweak vacuum survives the preheating stage for $t_\text{end} \lesssim t_\text{dec}$,
which yields the following upper bound for the Higgs-inflaton coupling:
\begin{claim}
\begin{align}
	c \lesssim \frac{\sqrt{6}}{8 \mu_\text{qtc}} \frac{m_\phi}{\Mpl} \ln \prn{ \frac{16 \pi^{\frac{3}{2}}}{3 \tilde \lambda} }
	\simeq 10^{-4} \  \com{\frac{0.1}{\mu_\text{qtc}}}\ \com{\frac{m_\phi}{10^{13} \,\text{GeV}}}.
	\label{eq:claim_qtc}
\end{align}
\end{claim}

\subsubsection*{Numerical simulation}
%%

%%%%%%%%%%%%%%%%
\begin{figure}[t]
\begin{minipage}{0.5\hsize}
\begin{center}
\includegraphics[scale = 0.8]{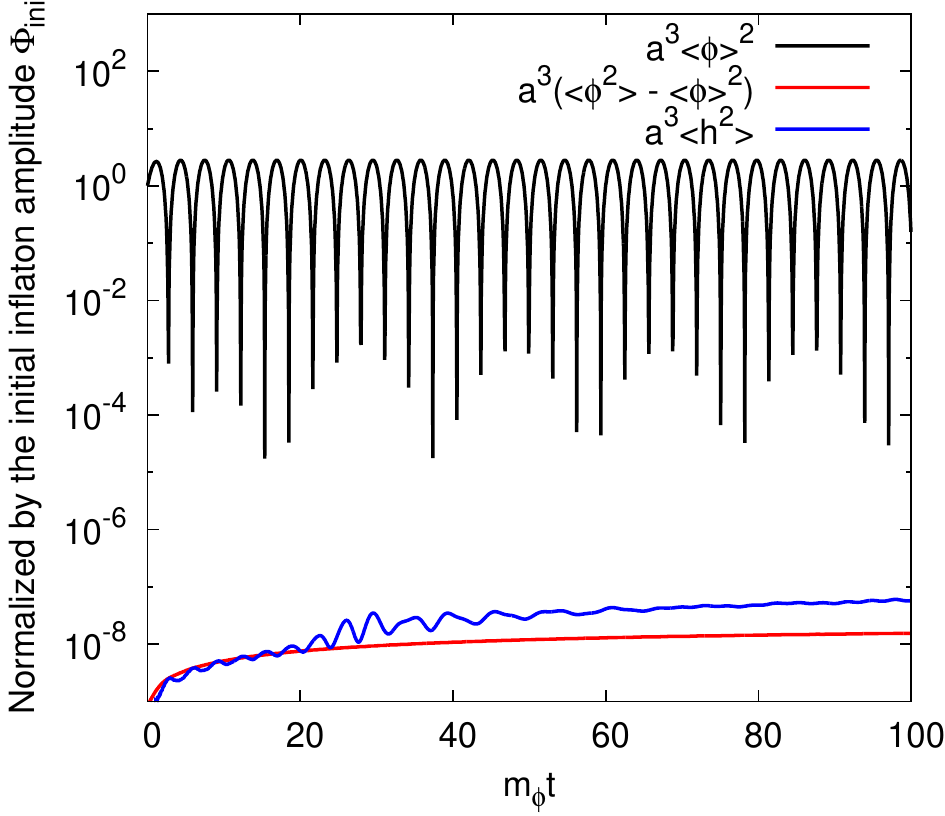}
\end{center}
\end{minipage}
\begin{minipage}{0.5\hsize}
\begin{center}
\includegraphics[scale = 0.8]{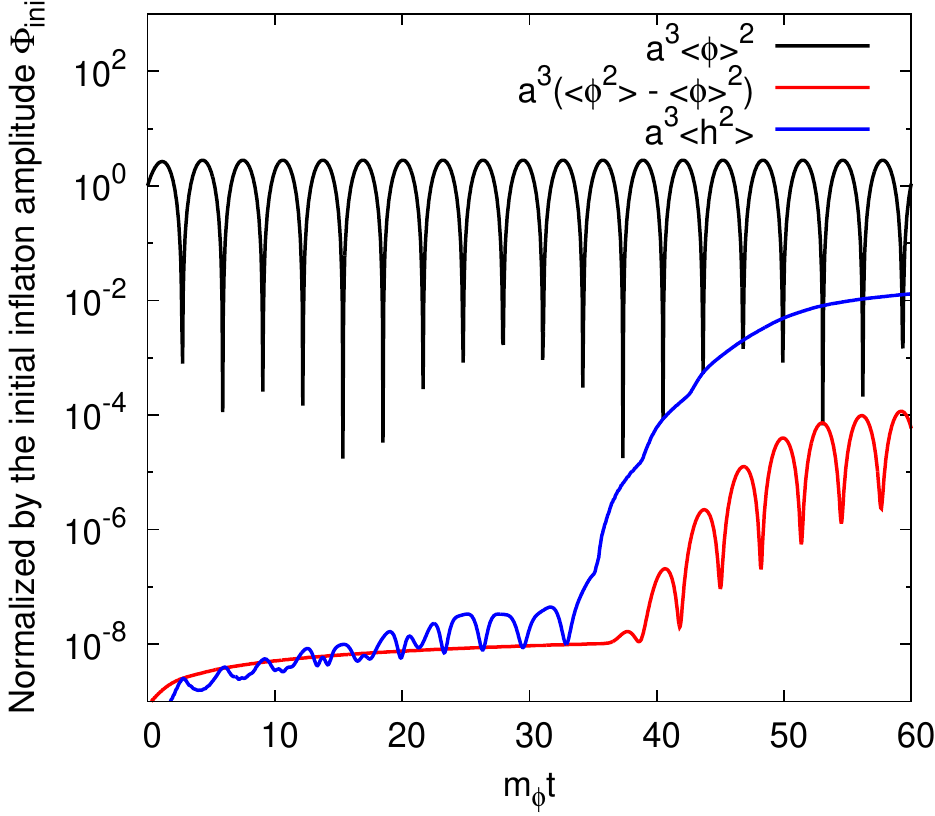}
\end{center}
\end{minipage}
\begin{minipage}{0.5\hsize}
\begin{center}
\includegraphics[scale = 0.8]{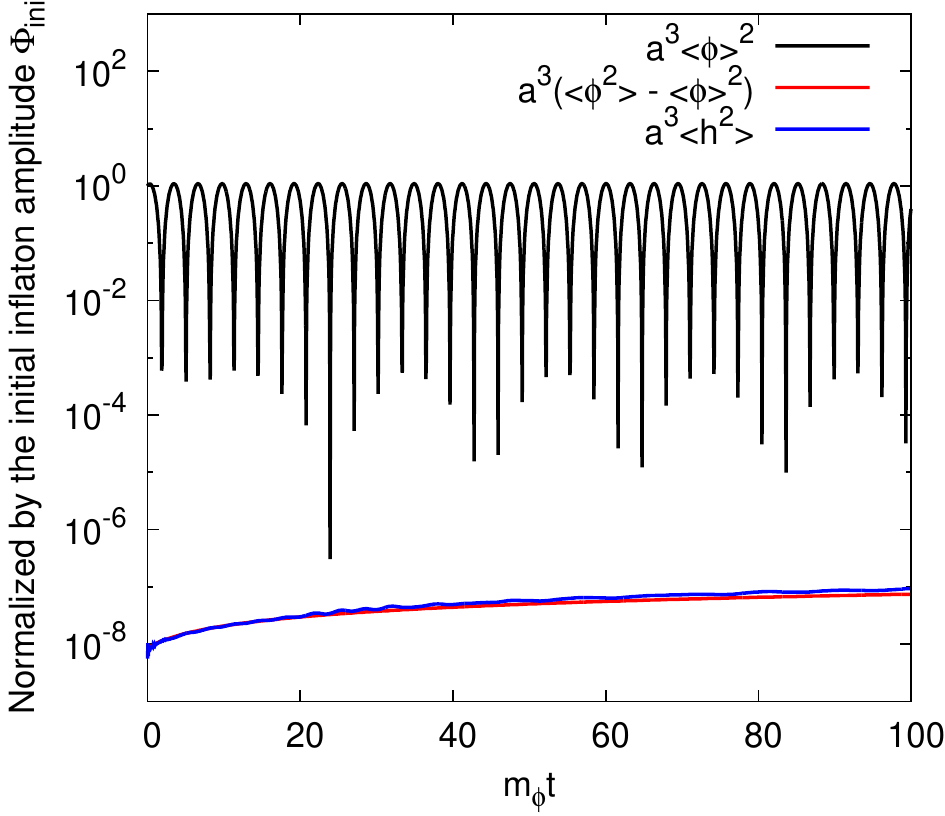}
\end{center}
\end{minipage}
\begin{minipage}{0.5\hsize}
\begin{center}
\includegraphics[scale = 0.8]{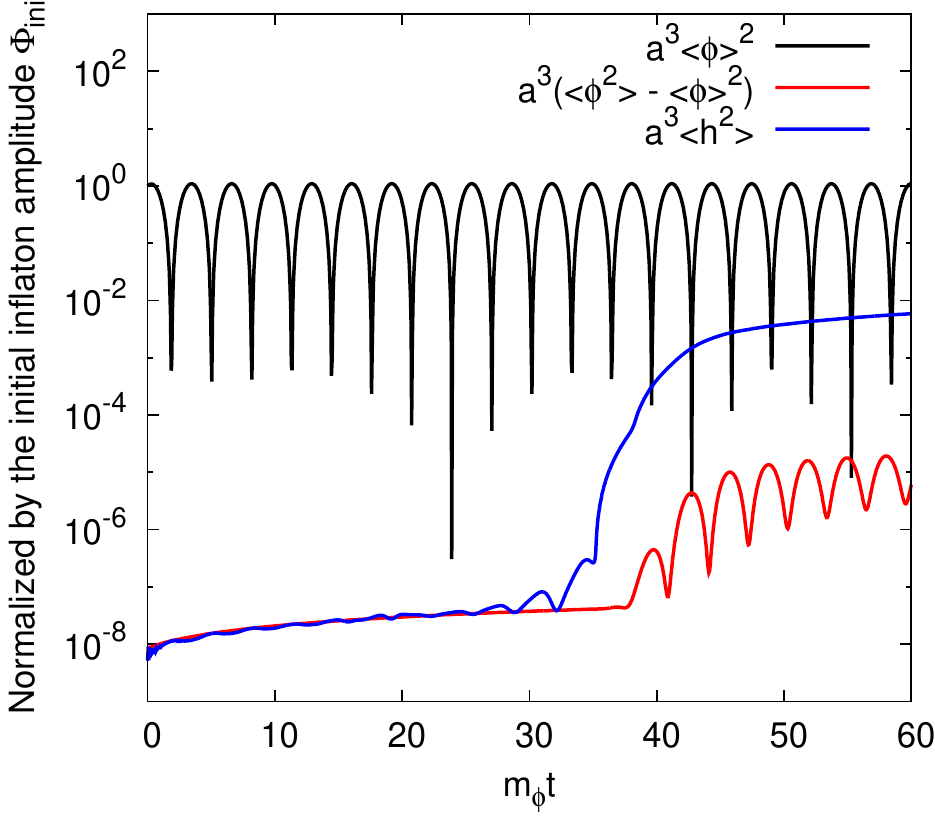}
\end{center}
\end{minipage}
\caption { \small
	Numerical calculation of the time evolution of the inflaton expectation value (black), 
	the inflaton dispersion (\textcolor{red}{red}) and the Higgs dispersion (\textcolor{blue}{blue}).
	We take the parameters as 
	$N = 128^{3}, \dd t = 10^{-3}/m_{\phi}, L = 10/m_\phi$
	and $m_{\phi} = 1.5\times 10^{13}\,\text{GeV}$.
	Upper left panel: $c = 1\times 10^{-4}$ and $\Phi_{\text{ini}} = \sqrt{2}\,\Mpl$.
	Upper right panel: $c = 2\times 10^{-4}$ and $\Phi_{\text{ini}} = \sqrt{2}\,\Mpl$.
	Lower left panel: $c = 1\times 10^{-4}$ and $\Phi_{\text{ini}} = \sqrt{0.2}\,\Mpl$.
	Lower right panel: $c = 2\times 10^{-4}$ and $\Phi_{\text{ini}} = \sqrt{0.2}\,\Mpl$.
	Higgs remains in the electroweak vacuum in the left panels, 
	while it rolls down to the true vacuum in the right panels.
}
\label{fig:qtc}
\end{figure}
%%%%%%%%%%%%%%%%

%%%%%%%%%%%%%%%%
\begin{figure}[t]
\begin{minipage}{0.5\hsize}
\begin{center}
\includegraphics[scale = 0.8]{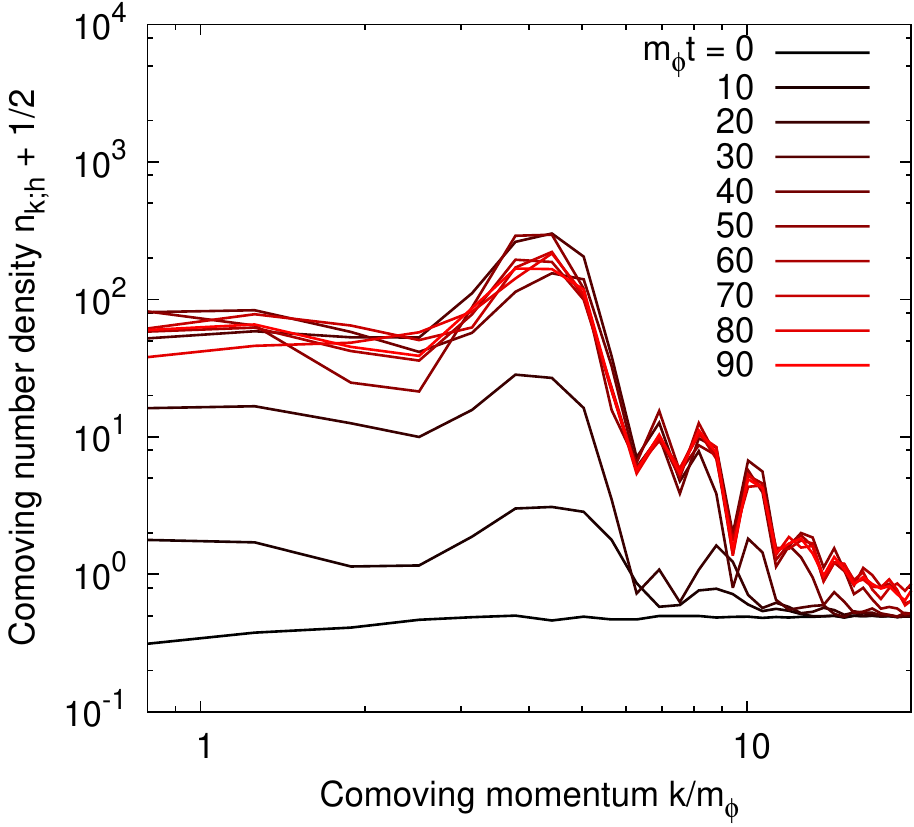}
\end{center}
\end{minipage}
\begin{minipage}{0.5\hsize}
\begin{center}
\includegraphics[scale = 0.8]{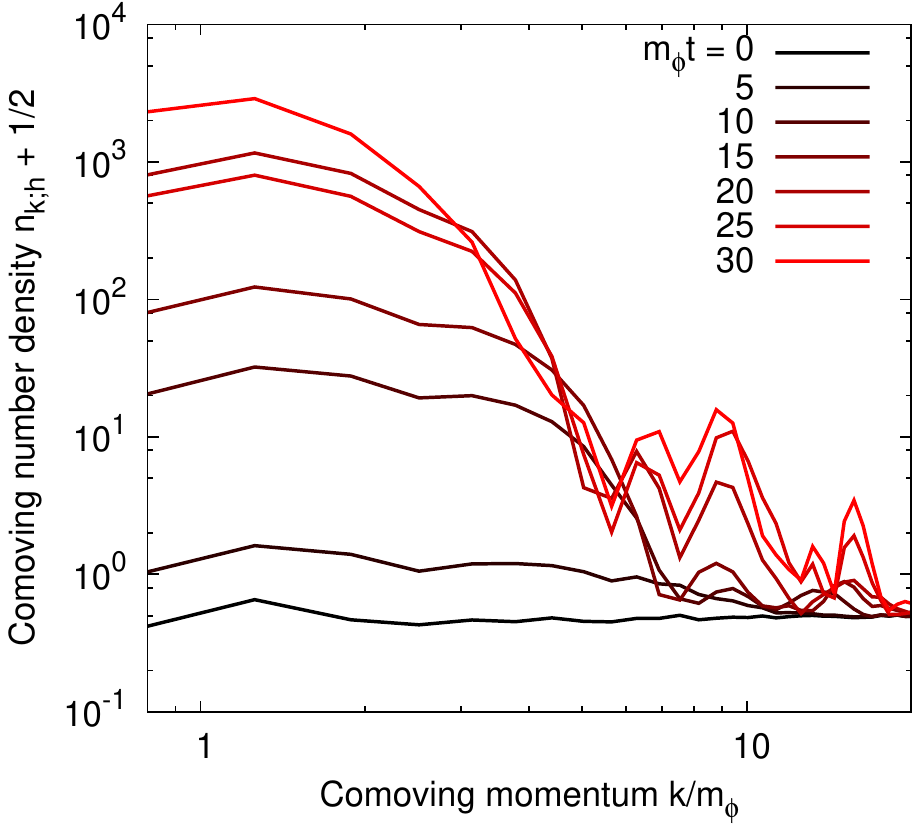}
\end{center}
\end{minipage}
\caption { \small
	Time evolution of the comoving number density of Higgs.
	We take the parameters as 
	$N = 128^{3}, \dd t = 10^{-3}/m_{\phi}, L = 10/m_{\phi}, 
	m_{\phi} = 1.5\times 10^{13}\,\text{GeV}$
	and $\Phi_{\text{ini}} = \sqrt{2}\,\Mpl$.
	Left panel: $c = 1\times 10^{-4}$.
	Right panel: $c = 2\times 10^{-4}$.
}
\label{fig:spe_qtc}
\end{figure}
%%%%%%%%%%%%%%%%

Below, we show the results of 3+1-dimensional classical lattice simulations for the quartic coupling case.
The main purpose here is to confirm the upper bound given in Eq.~\eqref{eq:claim_qtc}.

We solve the following classical equations of motion in the configuration space:
\begin{align}
	0 &= \ddot{\phi} + 3H\dot{\phi} - \frac{1}{a^2}\partial_i^2 \phi + \left( m_\phi^2 + c^2 h^2\right)\phi, \\
	0 &= \ddot{h} + 3H\dot{h} - \frac{1}{a^2}\partial_i^2 h + \left(c^2 \phi^2 + \lambda h^2\right)h,
\end{align}
for the scalar fields, and 
\begin{align}
	H^2 &= \frac{\langle \rho \rangle}{3\Mpl^2},
\end{align}
for the metric, where the total energy density is given by
\begin{align}
	\rho =& \frac{1}{2}\left(\dot{\phi}^2 + \frac{1}{a^2}\left(\partial_i \phi\right)^2 + m_\phi^2 \phi^2 \right) 
	+ \frac{1}{2}\left(\dot{h}^2 + \frac{1}{a^2}\left(\partial_i h\right)^2 \right) + \frac{\lambda}{4}h^4 + \frac{1}{2}c^2\phi^2h^2,
\end{align}
and $\langle ... \rangle$ denotes the spatial average.
We neglect fluctuations of the metric
because their effects are suppressed by the Planck mass.
The grid number is taken as $N = 128^3$ with a comoving edge size 
being $L = 10/m_\phi$ and
the time step is $\dd t = 10^{-3}/m_\phi$.
The inflaton mass is fixed as $m_\phi = 1.5\times 10^{13}\, \text{GeV}$.
We show results for $c = 1\times 10^{-4}$ and $c = 2\times 10^{-4}$.
Some details of our numerical calculation are summarized below:\footnote{
 	We have also confirmed our results by using {\tt LATTICEEASY}~\cite{Felder:2000hq}.
}
\begin{itemize}

\item We solve the classical equations of motion with the initial inflaton amplitude
$\Phi_{\text{ini}} = \sqrt{2}\,\Mpl$ or $\sqrt{0.2}\,\Mpl$.
By changing $\Phi_{\text{ini}}$, 
we test whether or not the inflation scale changes the bounds on the coupling strength.
We take the initial velocity as $\dot{\Phi}_{\text{ini}} = 0$,
but we have checked that our results are insensitive to this choice.
We also introduce gaussian initial fluctuations on the inflaton and Higgs field 
following Refs.~\cite{Polarski:1995jg, Khlebnikov:1996mc}, which mimic the quantum vacuum fluctuations.

\item
We renormalize masses of inflaton and Higgs originating from
the initial quantum fluctuations.
See App.~\ref{app:ren} for more details on this procedure.

\item 
There is another equation of motion for the metric sector, and it is redundant.
We have checked that our numerical calculation satisfies
the redundancy at least at $\mathcal{O}(10^{-3})$ precision.

\item In order to avoid numerical divergence, we add a sextic term to the Higgs potential.
We take the coefficient of the sextic term such that the Higgs field value $h_{\rm min}$
at the true minimum is $\tilde{\lambda}h_{\rm min}^2 = 5\times 10^{-8}\Mpl^2$.
The inequality $\tilde{\lambda}h_{\rm min}^2 \gg c^2\Phi^2$ is satisfied 
at the time when Higgs rolls down to the true vacuum,
and hence the sextic term does not affect the Higgs dynamics
before the electroweak vacuum decay. We have verified that it is not affected whether or not
the electroweak vacuum decays by the sextic term, by changing the coefficient of it.

\end{itemize}

In Fig.~\ref{fig:qtc}, we show the time evolution of the inflaton vacuum expectation value squared $\langle \phi \rangle^2$ (black), 
the inflaton dispersion $\langle \phi^2 \rangle - \langle \phi \rangle^2$ (\textcolor{red}{red})
and the Higgs dispersion $\langle h^2 \rangle$ (\textcolor{blue}{blue}).
Here $\langle \bullet^2 \rangle$ corresponds to $\vev{\com{\bullet, \bullet}_+} /2$,
and thus can be expressed by the statistical function, $G_{F; \bullet}$.
See Eq.~\eqref{eq:stat}.
They are multiplied by the scale factor to the third, where
the initial scale factor is taken as $a_{\text{ini}} = 1$. 
We take the coupling as $c = 1\times 10^{-4}$ for
the left panels and $c=2\times 10^{-4}$ for the right panels, respectively.
The initial inflaton amplitudes are $\Phi_{\text{ini}} = \sqrt{2}\,\Mpl$ 
for the upper panels and $\Phi_{\text{ini}} = \sqrt{0.2}\,\Mpl$ for the lower panels.

As it is clear from Fig.~\ref{fig:qtc}, Higgs stays at the electroweak vacuum for $c = 1\times 10^{-4}$,
while it rolls down to the true vacuum for $c = 2\times 10^{-4}$, independent of the initial inflaton amplitude.
It is consistent with our estimation~\eqref{eq:claim_qtc}.
In fact, an interesting feature of Eq.~\eqref{eq:claim_qtc} is that it does not depend
on the initial inflaton amplitude or the inflation scale as long as
the inequality~\eqref{eq:np_cond} is satisfied initially. 
This is because
the growth rate, $\mu_\text{qtc}$, does not much depend on the inflaton amplitude
for the broad resonance.\footnote{
As we will see later, the situation is completely different for the Higgs-curvature coupling case.
} 
Thus, the Higgs fluctuations are efficiently produced 
at the latest epoch [Eq.~\eqref{eq:qrt_end}] independent of the initial inflaton amplitude,
since the number of the inflaton oscillation is dominated by that epoch. 

We also plot the time evolution 
of the comoving number 
density of Higgs [Eq.~\eqref{eq:number_comoving}]
for $\Phi_{\text{ini}} = \sqrt{2}\,\Mpl$ 
in Fig.~\ref{fig:spe_qtc}.
The left panel shows the case with $c = 1\times 10^{-4}$ 
and the right panel does the case with $c = 2\times 10^{-4}$.
In the left panel, Higgs is efficiently produced at the beginning of the
oscillation, but the resonance shuts off due to the Hubble expansion.
After $t \simeq 30/m_{\phi}$, the comoving number density of Higgs remains almost
constant. On the other hand, in the right panel, the comoving number density of 
Higgs continues to grow resonantly.\footnote{
We do not plot the comoving number density of Higgs for $m_{\phi}t > 30$ in the right panel
because Higgs already rolls down to the true vacuum before that time.
} As a result, Higgs rolls down to the true vacuum once the condition~\eqref{eq:ema_cond} is satisfied.
Also, one can see that modes below $p_\ast$ is efficiently produced as expected.
The time evolution of the number density 
for $\Phi_{\text{ini}} = \sqrt{0.2}\,\Mpl$ is quite similar.

%%%%%%%%%%%%%%%%%%%%%%%%%%%%%%%%
%%%%%%%%%% Curvature Stabilization %%%%%%%%%%
%%%%%%%%%%%%%%%%%%%%%%%%%%%%%%%%
\subsection{Preheating via curvature stabilization: $\xi R h^2$}
\label{sec:curv}

\subsubsection*{Qualitative discussion}
Next, let us consider the Higgs-curvature coupling case.
In this case, the dispersion relation of Higgs is given by
\begin{align}
		\omega_{k;h}^2 (t) = \xi R 
	+ \frac{\bm{k}^2}{a^2(t)} 
	+ \delta m_{\text{self};h}^2 (t),
	\label{eq:disp_curv}
\end{align}
with the Ricci scalar being $R = 6 [ \ddot a /a + (\dot a/a)^2 ]$.
In the inflaton-oscillation dominated era,
the scale factor satisfies the following equalities:
\begin{align}
	\com{ \frac{\dot a (t)}{a (t)} }^2 = \frac{1}{6 \Mpl^2} \com{ \dot \phi^2 (t) + m_\phi^2 \phi^2 (t)},~~~
	\frac{\ddot a(t)}{a (t)} = \frac{1}{3 \Mpl^2} \com{ \frac{1}{2} m_\phi^2 \phi^2(t)  - \dot \phi^2 (t)}.
	\label{eq:scalefct}
\end{align}
Plugging Eqs.~\eqref{eq:sol_inf} and \eqref{eq:scalefct} into the dispersion relation,
we get
\begin{align}
	\omega_{k;h}^2 (t) + \Delta (t) \simeq   
	\frac{3}{2} \prn{\xi - \frac{1}{4}} \frac{m_\phi^2}{\Mpl^2} \Phi^2(t) \cos \prn{2 m_\phi t}
	+ \com{ \frac{\bm{k}^2}{a^2(t)} + \frac{\xi m_\phi^2}{2 \Mpl^2} \Phi^2 (t) }
	+ \delta m_{\text{self};h}^2 (t).
	\label{eq:disp_crv}
\end{align}
In contrast to the case of the quartic interaction [Eq.~\eqref{eq:disp_quart}],
the modes can be tachyonic in one oscillation when $\xi$ satisfies
$\xi < 3/16$ or $3/8 < \xi$.
Otherwise they are stable. If $\xi \Phi^2/\Mpl^2 < {\cal O}(1)$, 
the production becomes indistinguishable from the narrow resonance.\footnote{
In fact, it is almost stable for $\xi \Phi^2/\Mpl^2, \Phi^2/\Mpl^2 < 1$.
See Ref.~\cite{Tsujikawa:1999jh}.
} Thus, we concentrate on the case with\footnote{
We do not consider the case with $-\xi \gtrsim {\cal O}(1)$
because the Higgs-curvature coupling does not stabilize the electroweak vacuum during inflation.
}
\begin{align}
	\xi >  \com{\frac{\sqrt{2}\Mpl}{\Phi_\text{ini}}}^2,
	\label{eq:crv_lowb}
\end{align}
where an efficient particle production via the tachyonic instability occurs,
which we call tachyonic resonance.
Note that if this inequality holds, the Higgs stability during inflation is ensured since $\Phi_{\rm ini} \lesssim M_{\rm pl}$
leads to $\xi \gtrsim \mathcal O(1)$.\footnote{
	If $\Phi_\text{ini} \ll \Mpl$, it is possible to choose ${\cal O}(0.1) < \xi < (\Mpl/\Phi_{\rm ini})^2$.
	In such a case, Higgs is stable during inflation and also tachyonic resonance does not occur after inflation.
} In this case, the growth rate of the number density $X_k (t)$ is estimated as~\cite{Dufaux:2006ee}:\footnote{
	This is understood as follows. The time interval during which the Higgs with momentum $p=k/a$ becomes tachyonic in one inflaton oscillation is
	$\Delta t_k \sim m_\phi^{-1}(1-p^2/\xi R)$ for $p^2 \lesssim \xi R$ (here $R$ should be regarded as just a typical value).
	In one inflaton oscillation, these modes are enhanced as $\exp( \sqrt{\xi R} \Delta t_k ) \sim \exp(\sqrt{q}-p^2/p_*^{\rm (tac)2} )
	\sim \exp(\sqrt{q}-A_k/\sqrt{q})$, where $p_*^{\rm (tac)}$ is defined in (\ref{eq:ptac}).
}
\begin{align}
X_{k}(t) \simeq -\cfrac{x}{\sqrt{q}}\,A_{k} + 2x\sqrt{q},
\label{eq:growth_crv}
\end{align}
where
\begin{align}
q(t) = \cfrac{3}{4}\left(\xi - \cfrac{1}{4}\right) \cfrac{\Phi^2(t)}{\Mpl^2},~~~ 
A_{k}(t) = \cfrac{k^2}{a^2 (t)m_\phi^2} + \cfrac{\xi \Phi^2(t)}{2\Mpl^2}, 
\end{align}
with $x \simeq 0.85$. 
Recalling that $\Phi(t) \propto 1/t$,
one can see that the first few oscillations play the dominant role in the tachyonic resonance.
The growth rate is rather power law like,
contrary to the exponential growth in the previous case.
The typical physical momentum enhanced by the tachyonic resonance is given by
\begin{align}
p_\ast^{\text{(tac)}}(t) \equiv \frac{1}{\sqrt{x}}m_\phi q^{1/4}(t).  \label{eq:ptac}
\end{align}
In terms of $p_\ast^{\text{(tac)}}$, 
the condition for the efficient particle production via the tachyonic preheating
has the similar expression as Eq.~\eqref{eq:np_cond}:
\begin{align}
	p_\ast^{\text{(tac)}}(t) \gtrsim m_\phi.
	\label{eq:tac_cond}
\end{align}
The number density of Higgs field produced after the $j$-th passage of $\phi \sim 0$ is estimated as
\begin{align}
	n_h (t_j) &= \int_{\bm{k}/a(t_j)} n_{k;h} (t_j) \simeq \int_{\bm{k}/a (t_j)} e^{2 \sum_{i = 1}^{j}X_k(t_i)}
	\sim \frac{1}{16\pi^2} \sqrt{\frac{\pi}{2}} e^{n_{\text{eff}} (t_j)\,\mu_\text{crv} \sqrt{\xi} \frac{\Phi_\text{ini}}{\Mpl} } 
	\frac{a^3 (t_\text{ini})}{a^3 (t_j)} p_\ast^{\text{(tac)}^3}(t_\text{ini}),
\label{eq:number_crv}
\end{align}
where $\mu_\text{crv} \simeq 2x\sqrt{4/3}\simeq 2$ for $\xi \gtrsim {\cal O}(1)$, and $\Phi_{\text{ini}}$ is 
the initial inflaton amplitude.
The effective number of times of oscillation,
$n_\text{eff}(t_j) \equiv \sum_{i=1}^{j}\Phi(t_i)/\Phi_\text{ini}$, 
is a slightly time dependent function, which grows logarithmically.
We estimate $n_{\text{eff}} \simeq 1$ for $\Phi_{\text{ini}} \gtrsim \Mpl$.
This is because
the amplitude drastically decreases within the first oscillation, and hence
the first one or two oscillations dominate the Higgs production.
On the other hand, for $\Phi_{\text{ini}} \lesssim \Mpl$, the later oscillations can also be important
since the decrease of the amplitude is rather slow. For $\Phi_{\text{ini}} = \sqrt{0.2}\,\Mpl$, for example,
we roughly estimate $n_{\text{eff}} \simeq 1.5\mathchar`-2$.

Now we are in a position to derive the condition where the electroweak vacuum is stable during the preheating.
Contrary to the quartic stabilization,
the curvature coupling becomes tachyonic every crossing around $\phi \sim 0$.
Its time scale is of the order of $m_\phi^{-1}$.
During that period, the Higgs field grows towards an unwanted deeper minimum
owing to the tachyonic mass terms from the curvature coupling $|\xi R|_{\phi \sim 0} \sim q m_\phi^2$
or the Higgs self coupling $|\delta m_\text{self;h}^2|$.
While the curvature coupling dominates the tachyonic mass,
the growth rate of the Higgs field is estimated as $\sqrt{|\xi R|}_{\phi \sim 0} \Delta t \sim \sqrt{q}$.
This is nothing but the efficiency of the resonance.
Once that from the Higgs four point couplings becomes larger, the growth rate of the Higgs field is accelerated;
$|\delta m_\text{self;h}^2| \gtrsim q m_\phi^2$.
Almost at the same time, the tachyonic mass term from the Higgs self coupling, $\delta m_\text{self;h}^2$, 
exceeds the positive effective mass
from the curvature coupling, $\xi R_{\phi \sim \Phi} \sim q m_\phi^2$, and the vacuum decay is triggered.
Thus, we estimate the condition where Higgs is stable against 
the tachyonic mass as
\begin{align}
	\abs{ \delta m_{\text{self}; h}^2 (t_j) }_{\xi R \sim 0} \lesssim q m_\phi^2\leftrightarrow
	\frac{3 \tilde \lambda}{16 \pi^2} \sqrt{\frac{\pi}{2q (t_\text{ini})}} \frac{a (t_\text{end})}{a (t_\text{ini})} \,
	e^{n_{\text{eff}} (t_\text{end})\,\mu_\text{crv} \sqrt{\xi} \frac{\Phi_\text{ini}}{\Mpl} } \lesssim 1.
	\label{eq:ema_cond_2}
\end{align}
Again, one can see that not only the long wave length mode but all the modes below $q^{1/2} m_\phi$
grows towards the lower potential energy regime, overcoming the pressure of spatial gradients,
if this condition is violated.
As a result, the electroweak vacuum survives the tachyonic resonance for
\begin{claim} 
\begin{align}
		\xi \lesssim \frac{1}{n_{\text{eff}}^2\,\mu_\text{crv}^2} 
		\com{\frac{\Mpl}{\Phi_\text{ini}}}^2 \com{ \ln \prn{\frac{16 \pi^2}{3 \tilde \lambda} \sqrt{\frac{2}{\pi}}} }^2
		\simeq 10 \ \com{\frac{2}{n_{\text{eff}}\,\mu_\text{crv}}}^2 \com{\frac{\sqrt{2}\,\Mpl}{\Phi_\text{ini}}}^2,
	\label{eq:claim_crv}
\end{align}
\end{claim}
\noindent
where we have taken $q\simeq 1$ in the logarithm and
neglect the scale factor dependence.
This is because $q$ cannot be much larger than $\mathcal O(1)$ in the present case
since otherwise the tachyonic growth is catastrophic.

Here let us clarify differences of our analysis from that of Ref.~\cite{Herranen:2015ima},
in which a similar study is performed for the Higgs-curvature coupling case.
There are two main differences.
First, the criteria for the Higgs field to escape from the electroweak vacuum is different.
The authors in Ref.~\cite{Herranen:2015ima}
compare the Higgs dispersion for the over-horizon modes
$\sqrt{\langle h^2\rangle_{aH}}$ and $h_{\rm max}$ ($\Lambda_I$ in their notation).
Instead, we take into account the stabilization effect from the Higgs-curvature coupling.
We compare the effective masses induced 
by the Higgs self coupling and the Higgs-curvature coupling.
Thus, the bound we obtain here is a bit weaker than their result.
For definiteness, we perform a classical lattice simulation and confirm our estimation in the next section.
Second, the treatment of the Higgs production caused 
by the Higgs-curvature coupling is also different.
They treated the particle production caused by
a transition of the mean-value of 
the Ricci scalar $R$ after inflation~\cite{Ford:1986sy} and
the oscillation of $R$ around the mean-value separately as a rough approximation.
In our analysis, these effects are simultaneously included 
in the classical lattice simulation as we shall see below.

\subsubsection*{Numerical simulation}
%%

%%%%%%%%%%%%%%%%
\begin{figure}[t]
\begin{minipage}{0.5\hsize}
\begin{center}
\includegraphics[scale = 0.8]{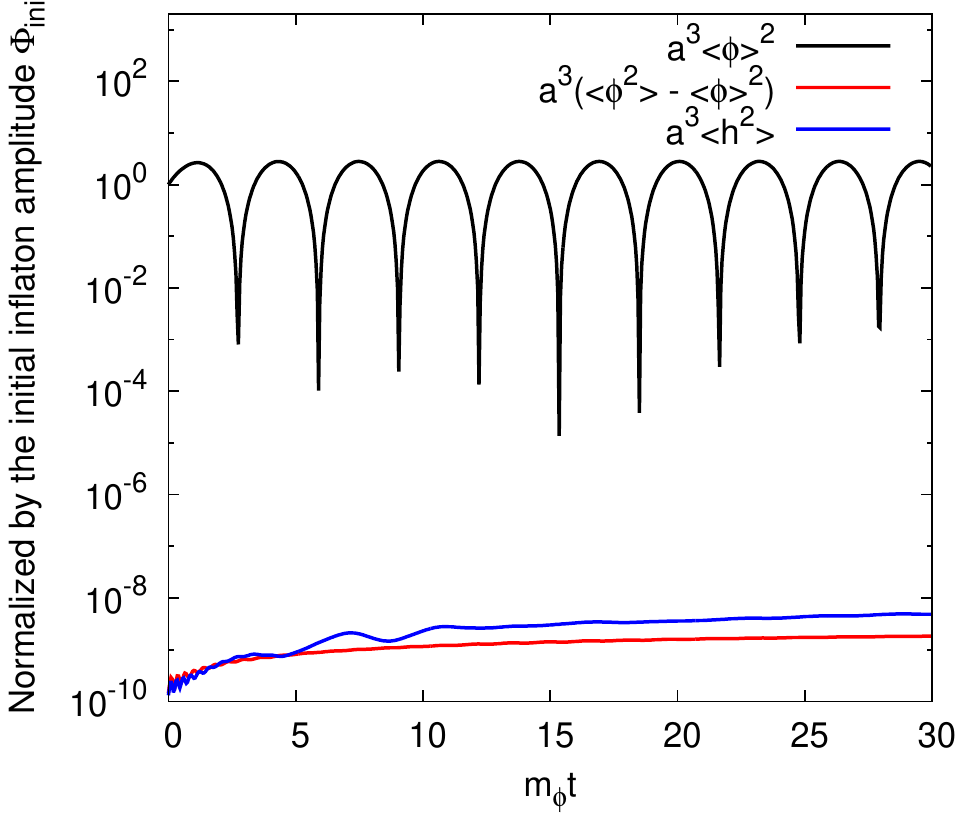}
\end{center}
\end{minipage}
\begin{minipage}{0.5\hsize}
\begin{center}
\includegraphics[scale = 0.8]{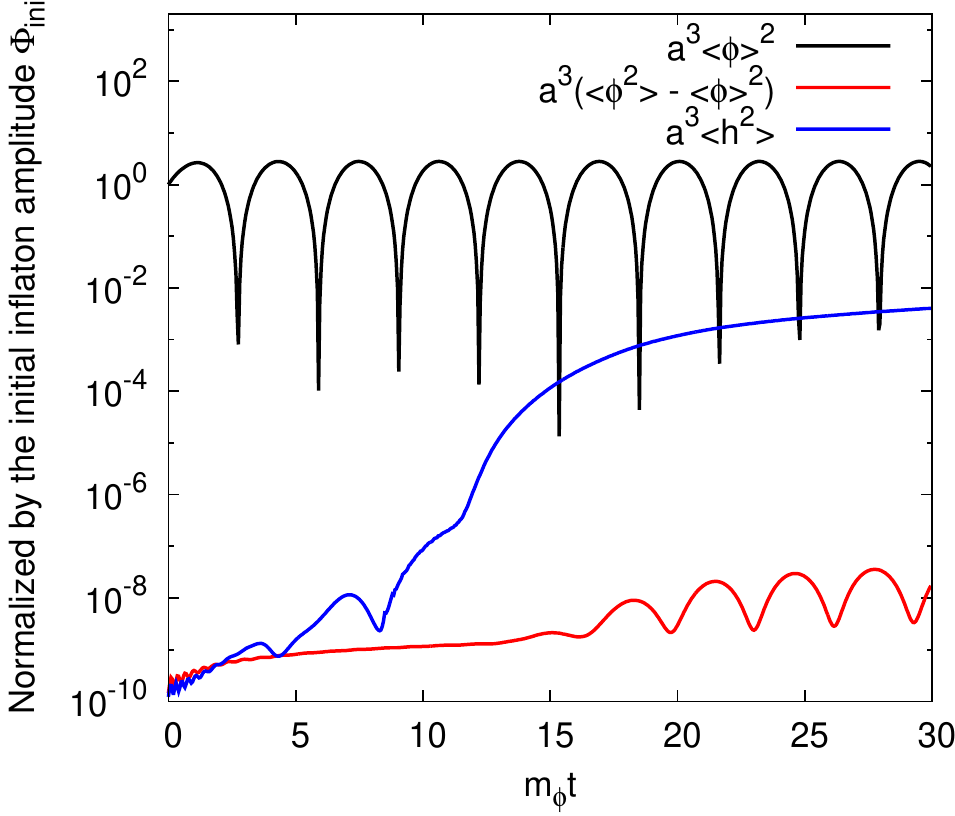}
\end{center}
\end{minipage}
\begin{minipage}{0.5\hsize}
\begin{center}
\includegraphics[scale = 0.8]{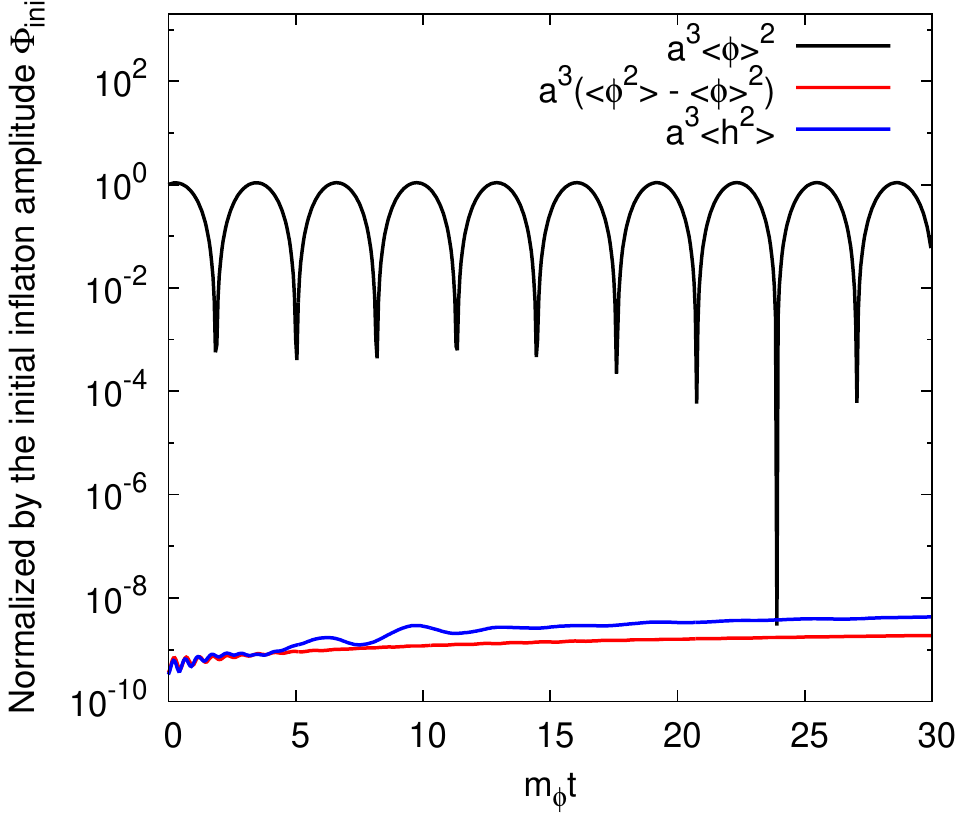}
\end{center}
\end{minipage}
\begin{minipage}{0.5\hsize}
\begin{center}
\includegraphics[scale = 0.8]{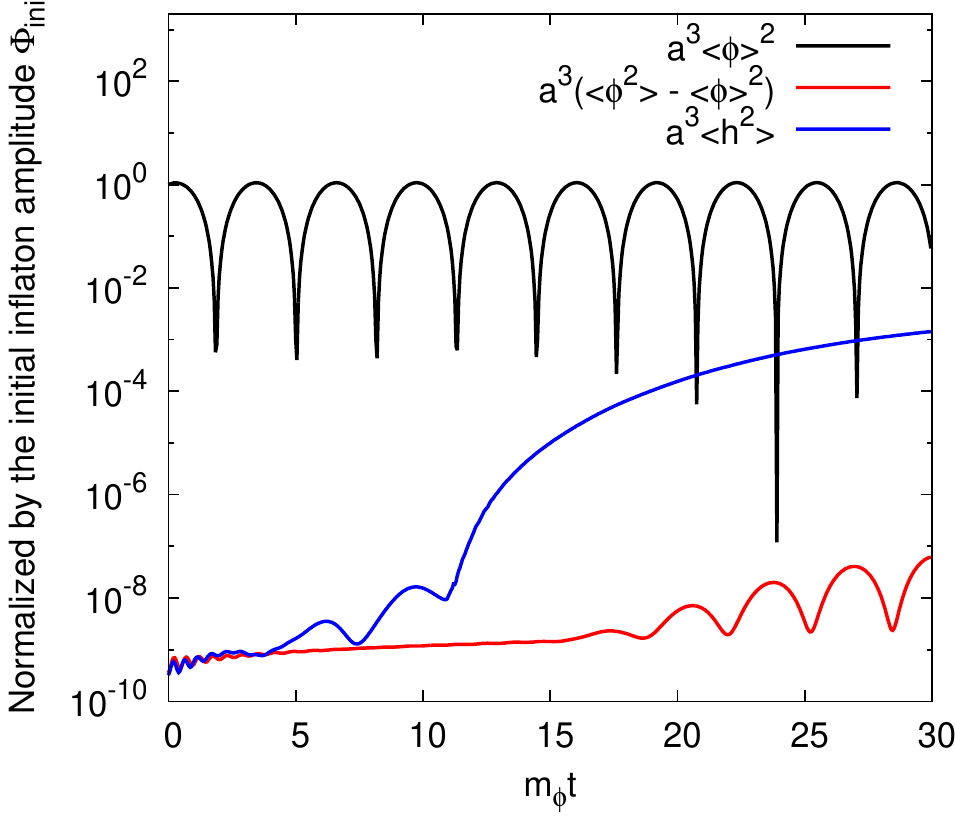}
\end{center}
\end{minipage}
\caption { \small
	Numerical calculation of the time evolution of the inflaton expectation value (black), 
	the inflaton dispersion (\textcolor{red}{red}) and the Higgs dispersion (\textcolor{blue}{blue}).
	We take the parameters as 
	$N = 128^{3}, \dd t = 10^{-3}/m_{\phi}$ and
	$m_{\phi} = 1.5\times 10^{13}\,\text{GeV}$.
	The comoving edge size is
	$L = 20/m_{\phi}$ for $\Phi_{\text{ini}} = \sqrt{2}\,\Mpl$
	and $L = 40/m_{\phi}$ for $\Phi_{\text{ini}} = \sqrt{0.2}\,\Mpl$.
	Upper left panel: $\xi = 10$ and $\Phi_{\text{ini}} = \sqrt{2}\,\Mpl$.
	Upper right panel: $\xi = 20$ and $\Phi_{\text{ini}} = \sqrt{2}\,\Mpl$.
	Lower left panel: $\xi = 20$ and $\Phi_{\text{ini}} = \sqrt{0.2}\,\Mpl$.
	Lower right panel: $\xi = 30$ and $\Phi_{\text{ini}} = \sqrt{0.2}\,\Mpl$.
	Higgs remains in the electroweak vacuum in the left panels, 
	while it rolls down to the true vacuum in the right panels.
}
\label{fig:crv}
\end{figure}
%%%%%%%%%%%%%%%%

%%%%%%%%%%%%%%%%
\begin{figure}[t]
\begin{minipage}{0.5\hsize}
\begin{center}
\includegraphics[scale = 0.8]{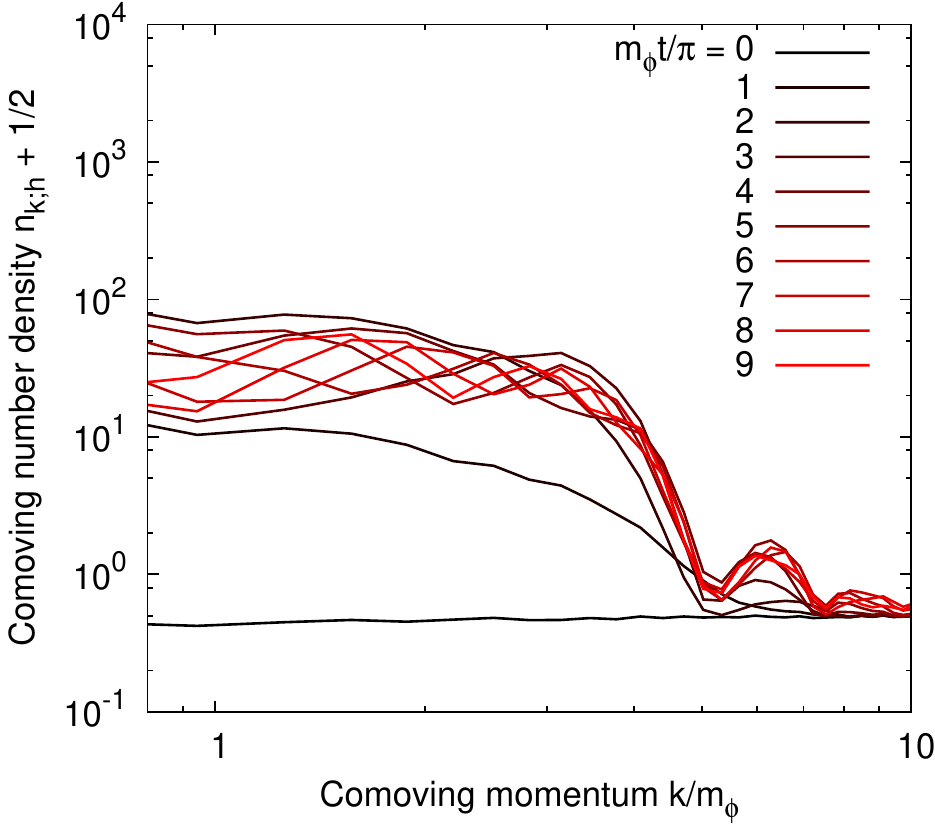}
\end{center}
\end{minipage}
\begin{minipage}{0.5\hsize}
\begin{center}
\includegraphics[scale = 0.8]{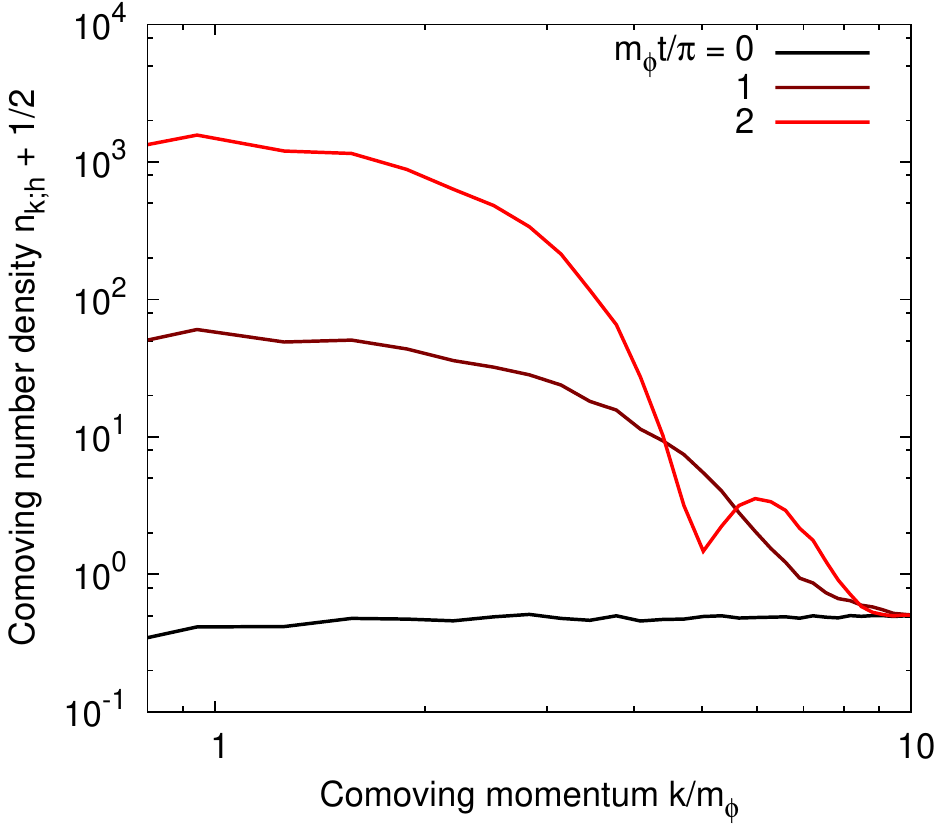}
\end{center}
\end{minipage}
\caption { \small
	Time evolution of the comoving number density of Higgs.
	We take the parameters as 
	$N = 128^{3}, \dd t = 10^{-3}/m_{\phi}, L = 20/m_{\phi}, 
	m_{\phi} = 1.5\times 10^{13}\,\text{GeV}$
	and $\Phi_{\text{ini}} = \sqrt{2}\,\Mpl$.
	Left panel: $\xi = 10$.
	Right panel: $\xi = 20$.
	We evaluate the number density at the end points of the oscillations.
	We show the number density only before Higgs rolls down to the true vacuum 
	for the right panel.
}
\label{fig:spe_crv}
\end{figure}
%%%%%%%%%%%%%%%%

In the following, we show the results of classical lattice simulations for the curvature coupling case.
The main purpose of this sub-section is to confirm the upper bound given in Eq.~\eqref{eq:claim_crv}.

We solve the following classical equations of motion
in the configuration space:
\begin{align}
	0 &= \ddot{\phi} + 3H\dot{\phi} - \frac{1}{a^2}\partial_i^2 \phi + 2\frac{\xi}{\Mpl^2}
	\left[\dot{\phi}\dot{h} - \frac{1}{a^2}\partial_i \phi \partial_i h\right]h + \left(1+\frac{\xi h^2}{\Mpl^2}\right)m_\phi^2 \phi, \\
	0 &= \ddot{h} + 3H\dot{h} - \frac{1}{a^2}\partial_i^2 h + \frac{\xi}{\Mpl^2}
	\left[2m_\phi^2\phi^2 - \dot{\phi}^2 + \frac{1}{a^2}\left(\partial_i \phi\right)^2 \right]h + \lambda h^3,
\end{align}
for the scalar fields, and
\begin{align}
	H^2 = \frac{\langle \rho \rangle}{3\Mpl^2},
\end{align}
for the metric, where the energy density is given by
\begin{align}
	\rho &= \frac{1}{2}\left(1 + \frac{\xi h^2}{\Mpl^2}\right)\left(\dot{\phi}^2 + \frac{1}{a^2}\left(\partial_i \phi\right)^2\right)
	+ \frac{1}{2}\left(1 + 2\frac{\xi h^2}{\Mpl^2}\right)m_\phi^2 \phi^2 
	+ \frac{1}{2}\left(\dot{h}^2 + \frac{1}{a^2}\left(\partial_i h\right)^2\right) + \frac{\lambda}{4}h^4.
\end{align}
Namely, we solve the equations in the Einstein frame.
We  only keep up to first-order terms 
in $\xi h^2/\Mpl^2$ and $\xi^2 h^2/\Mpl^2$. This treatment
is justified because $\xi h^2/\Mpl^2, \xi^2 h^2/\Mpl^2 \ll 1$
always holds in our numerical calculation.
We take the grid number as $N = 128^3$ 
with a comoving edge size being $L = 20/m_{\phi}$ for 
the initial inflaton amplitude $\Phi_{\text{ini}} = \sqrt{2}\,\Mpl$
and $L = 40/m_{\phi}$ for $\Phi_{\text{ini}} = \sqrt{0.2}\,\Mpl$\footnote{
	The typical momentum of our interest is smaller than 
	that in the quartic coupling case. This is why we take 
	the comoving edge size smaller than that in the quartic coupling case here.
}
and the time step as $\dd t = 10^{-3}/m_\phi$.
We fix the inflaton mass as $m_\phi = 1.5\times 10^{13}\, \text{GeV}$.
Some details of our numerical calculation are summarized below
(the same as those in the quartic coupling case):
\begin{itemize}

\item 
We start to solve the classical equations of motion with
$\Phi_{\text{ini}} = \sqrt{2}\,\Mpl$ or $\sqrt{0.2}\,\Mpl$.
We set the initial velocity as 
$\dot{\Phi}_{\text{ini}} = 0$.
We introduce gaussian initial fluctuations in the inflaton and Higgs fields
which arise from the quantum fluctuations.

\item
We renormalize masses of inflaton and Higgs originating from
the initial quantum fluctuations.
See App.~\ref{app:ren} for more details on this procedure.

\item
We use the redundancy of the equations of motion for the metric 
as a check of our numerical calculation,
and verified that our numerical calculation satisfies
the redundancy at least at $\mathcal{O}(10^{-3})$ precision.

\item We add a sextic term to the Higgs potential to stabilize it in our calculation.
We take the coefficient such that the Higgs field value at the true minimum is $\tilde{\lambda}h_{\rm min}^2 = 5\times 10^{-8}\Mpl^2$.

\end{itemize}

In Fig.~\ref{fig:crv}, we show the time evolution of 
the inflaton vacuum expectation value squared 
$\langle \phi \rangle^2$ (black), the inflaton dispersion 
$\langle \phi^2 \rangle - \langle \phi \rangle^2$ (\textcolor{red}{red})
and the Higgs dispersion $\langle h^2 \rangle$ (\textcolor{blue}{blue}),
which are multiplied by the scale factor to the third whose initial value is $a_{\text{ini}} = 1$.
Again, note that the dispersions are related with the statistical function.
We take the Higgs-curvature coupling and the initial inflaton amplitude as follows:
$\xi = 10$ and $\Phi_{\text{ini}} = \sqrt{2}\,\Mpl$ for the upper left panel,
$\xi = 20$ and $\Phi_{\text{ini}} = \sqrt{2}\,\Mpl$ for the upper right panel,
$\xi = 20$ and $\Phi_{\text{ini}} = \sqrt{0.2}\,\Mpl$ for the lower left panel and
$\xi = 30$ and $\Phi_{\text{ini}} = \sqrt{0.2}\,\Mpl$ for the lower right panel, respectively.

For the $\Phi_{\text{ini}} = \sqrt{2}\,\Mpl$ cases (the upper panels),
the Higgs field remains in the electroweak vacuum
for $\xi = 10$, while it rolls down to the true vacuum for $\xi = 20$. 
Thus, the condition~\eqref{eq:claim_crv} is consistent with our numerical calculation
within a factor of two.\footnote{
The difference between Eq.~\eqref{eq:claim_crv} and our numerical 
calculation may be due to the fact that the inflaton amplitude decreases drastically
within the first oscillation for $\Phi_{\text{ini}} = \sqrt{2}\,\Mpl$, while we assume 
that the amplitude is constant to derive Eq.~\eqref{eq:claim_crv}.
Our main purpose is, however, an order estimation of the critical value of
the curvature coupling, and hence Eq.~\eqref{eq:claim_crv} is enough.
}
On the contrary to the quartic coupling case, the critical value of the Higgs-curvature coupling 
depends on $\Phi_\text{ini}$. 
Indeed, for the $\Phi_{\text{ini}} = \sqrt{0.2}\,\Mpl$ cases (the lower panels),
Higgs remains in the electroweak vacuum for $\xi = 20$, 
while it rolls down to the true vacuum for $\xi = 30$.
Again, it is consistent with our estimation~\eqref{eq:claim_crv} once we include 
the effect of $n_{\text{eff}} \simeq 1.5\mathchar`-2$.

In Fig.~\ref{fig:spe_crv}, we also plot the time evolution of 
the comoving number density of Higgs [Eq.~\eqref{eq:number_comoving}]
for $\Phi_{\text{ini}} = \sqrt{2}\,\Mpl$.
The curvature coupling is $\xi = 10$ for the left panel and 
$\xi = 20$ for the right panel, respectively.
We have evaluated the number density at the end points of the oscillations 
since it is well-defined only at around these points for the curvature coupling case.
Higgs is created dominantly within the first few oscillations.
This is because the growth rate depends on the inflaton amplitude.
The time evolution of the number density 
for $\Phi_{\text{ini}} = \sqrt{0.2}\,\Mpl$ are quite similar.

%%%%%%%%%%%%%%%%%%%%%%%%%%%%%%%%
%%%%%%% Role of Higgs-Radiation Coupling %%%%%%%
%%%%%%%%%%%%%%%%%%%%%%%%%%%%%%%%
\section{Higgs-Radiation Coupling}
\label{sec:higgs_radiation}

In the previous section,
we have neglected the interactions between Higgs and radiation 
via Yukawa and gauge couplings
to illustrate the impacts of Higgs-inflaton coupling.
In this section, we investigate whether finite density corrections from the Higgs-radiation coupling
could relax the bounds~\eqref{eq:claim_qtc} and \eqref{eq:claim_crv}.
In contrast to the vacuum corrections, which destabilize the Higgs potential via fermion loops ({\it i.e.}~top quark),
finite density corrections tend to stabilize the Higgs field to its enhanced symmetry point once
top quarks and/or electroweak gauge bosons are produced.
There are three ways to produce these particles in the course of reheating dynamics:
\begin{description}
	\item[{\bf Instant preheating:}] For a relatively large Higgs-inflaton coupling,
	Higgs produced non-perturbatively at $\phi \sim 0$ decays into top quarks
	within one oscillation at a large field value of inflaton
	where Higgs becomes heavy~\cite{Felder:1998vq}. 
	If this decay is prompt, the efficiency of 
	the broad/tachyonic resonance is reduced and the decay products stabilize the 
	Higgs at its enhanced symmetry point.
	
	\item[{\bf Annihilation:}] If the number density of Higgs becomes large owing to the broad/tachyonic resonance,
	the annihilation of Higgs into top quarks and electroweak gauge bosons becomes significant~\cite{Bezrukov:2008ut,Moroi:2013tea}.
	This reduces the efficiency of resonance. 
	Also produced top quarks and gauge bosons may stabilize the Higgs
	at its enhanced symmetry point.
	
	\item[{\bf Complete reheating:}]
	The Higgs-inflaton/-curvature coupling alone cannot lead to a complete decay of inflaton,
	and hence an additional interaction of inflaton which completes the reheating is required.
	Since the radiation is produced before the complete decay of inflaton 
	via this interaction ~\cite{Giudice:2000ex,Harigaya:2013vwa,Mukaida:2015ria},
	the abundant top quarks and electroweak gauge bosons may stabilize the Higgs at its enhanced symmetry point.
\end{description}
We discuss first two effects in this section.
To avoid complications,
effect of the complete reheating is explained in Sec.~\ref{sec:reheating}
because it depends on another unknown parameter, \textit{i.e.} reheating temperature.

%%%%%%%%%%%%%%%%%%%%%%%%%%%%%%%%
%%%%%%%%%% Instant Preheating %%%%%%%%%%
%%%%%%%%%%%%%%%%%%%%%%%%%%%%%%%%
\subsection{Instant preheating}
\label{sec:inst_preheating}
Since Higgs couples with top quarks via the sizable Yukawa coupling,
it decays into top quarks at a large field value region of inflaton.
The decay might affect 
the preheating dynamics in two ways:
(i) reduce the efficiency of Higgs production,
(ii) stabilize the Higgs potential by the screening mass term from decay products.
We first give an overview of these effects below.

At the early stage of the preheating, 
the dynamics depends on the stabilization mechanism.
For the quartic stabilization $c^2 \phi^2 h^2$,
the efficiency of the resonance is independent of the coupling $c$.
Hence, the Higgs decay, which is proportional to $c$,
dominates over the Higgs resonant production at the early epoch.
Radiation is effectively produced via the Higgs decay at this epoch,
and it may stabilize the Higgs potential. It corresponds to the effect (ii).
For the curvature stabilization $\xi R h^2$, on the contrary,
both the efficiency of the resonance and the decay is proportional to $\xi$.
Thus, the resonant Higgs production always dominates over the Higgs decay,
and the effect (ii) is negligible.
At the late stage, the Higgs resonant production dominates over the Higgs decay
even in the case of the quartic stabilization.
Hence, the Higgs decay just reduces the efficiency of the Higgs resonant production 
for both the quartic/curvature stabilization, which corresponds to the effect (i).

From now we discuss the effects (i) and (ii) in detail.
We first consider the effect (i), neglecting the effect (ii).
Then we discuss the effect (ii) for the quartic coupling case.
We will see that it is relevant only if the Higgs-inflaton coupling is quite large,
say $c \gtrsim {\cal O}(10^{-1})$.

\subsubsection*{Slow decay of Higgs [Late stage]}
First, let us concentrate on the first effect (i).
Thus, we neglect the second effect (ii),
assuming that the resonance is not terminated by the back-reaction of decay products.
We discuss the quartic stabilization in detail.
A similar discussion holds for the curvature stabilization.

We estimate a typical decay rate of Higgs boson by taking an oscillation average,
which gives $\bar \Gamma_{h \to t \bar t} = (3 \alpha_t /2) \bar m_{H;h} \sim (3 \alpha_t / 2 \sqrt{2}) c \Phi$.
Here the bar indicates the oscillation average, $\Gamma_{h \to t \bar t}$ is the decay rate of Higgs into top quarks,
$m_{H; h}$ is the effective mass of Higgs, 
and $ \alpha_t \equiv y_t^2 / (4 \pi) \sim 0.02$
is the top Yukawa coupling at $m_\phi \sim 10^{13}\,\text{GeV}$. 
The decay reduces the growth of Higgs fluctuations as $n_h \propto e^{2 \mu m_\phi t - \bar \Gamma_{h \to t \bar t} t}$.
As a result, the decay time given in Eq.~\eqref{eq:qrt_dec} becomes slightly longer:
\begin{align}
	m_\phi t_\text{dec} \sim \frac{1}{2 \mu_{\rm qtc}} \ln \prn{ \frac{16 \pi^\frac{3}{2}}{3 \tilde \lambda} }
	+ \frac{\sqrt{3} \alpha_t}{2 \mu_{\rm qtc}} \ \frac{ c \Mpl}{m_\phi}.
\end{align}
By comparing it with $t_\text{end}$ given in Eq.~\eqref{eq:qrt_end},
we estimate the impacts of the Higgs decay on the upper bound [Eq.~\eqref{eq:claim_qtc}]
as follows:
\begin{align}
	c \lesssim 
	10^{-4} \  \com{\frac{0.1}{\mu_{\rm qtc}}}\ \com{\frac{m_\phi}{10^{13} \,\text{GeV}}}
	\com{ 1 - 0.1 \prn{ \frac{\alpha_t}{0.02}} \prn{\frac{0.1}{\mu_{\rm qtc}}} }^{-1}.
\end{align}
Similarly, in the case of the curvature stabilization,
we find
\begin{align}
	\xi
	\lesssim 10 \times \com{\frac{2}{n_{\text{eff}}\,\mu_{\rm crv}}}^2 \com{\frac{\sqrt{2}\,\Mpl}{\Phi_\text{ini}}}^2
	\com{1 - 0.04 \prn{\frac{\alpha_t}{0.02}}\prn{\frac{2}{\mu_{\rm crv}}}}^{-2}.
\end{align}
Even if we optimistically estimate the decay rate of Higgs,
the effect (i) does not change the upper bound by a large amount,
within uncertainties of our order of magnitude estimation.
Note here that the decay of Higgs fluctuations into top quarks may be much suppressed in reality.
This is because the typical life time of Higgs is prolonged for Higgs fluctuations are relativistic,
and because the top quarks acquire a finite density mass correction from the Higgs fluctuations,
which could suppress the Higgs decay into two top quasi-particles kinematically.
Since the purpose of this subsection is to confirm that
our estimation is not affected even if we take into account the effect (i),
we do not further investigate this effect.

\subsubsection*{Instant decay of Higgs [Early stage]}
The next step is to include the second effect (ii).
As we saw at the beginning of this sub-section,
the Higgs decay dominates over the Higgs resonant production
at the early epoch of the preheating stage for the quartic coupling case.
In fact, for a sizable coupling $c$,
the resonance can be killed by the back-reaction of decay products of Higgs.

If the quartic coupling $c$ satisfies $\bar \Gamma_{h \to t \bar t} \gg m_\phi / \pi \leftrightarrow \pi \alpha_t c \Phi \gg m_\phi$, 
Higgs produced at $\phi \sim 0$ decays completely before the inflaton moves back to its potential origin $\phi \sim 0$.
While this condition holds, the inflaton decays with the following rate~\cite{Mukaida:2012qn,Mukaida:2012bz}:\footnote{
Here note that once the condition, $\pi \alpha_t c \Phi (t) \gg m_\phi$, is violated owing to the reduction of inflaton amplitude, \textit{e.g.}~by the cosmic expansion,
the effective decay rate of inflaton becomes $\sim \Gamma_\text{inst} \times [  (\sqrt{3}  \pi / 2) (\alpha_t c \Phi / m_\phi)^{3/2}]$,
which decreases faster than the Hubble parameter.
}
\begin{align}
	\Gamma_\text{inst} \sim \frac{c^2}{4 \pi^4 [3\alpha_t /2 ]^\frac{1}{2}}\ m_\phi~~~
	\text{for}~~~\pi \alpha_t c \Phi (t) \gg m_\phi.
\end{align}
Contrary to the previous case,
one can show that the Higgs becomes non-relativistic at its decay.
Also, the effective mass of top quarks from the Higgs fluctuations can be safely neglected
because the effective mass of Higgs from the inflaton field at its decay is sufficiently large in this case.
Assuming that the conditions, $\pi \alpha_t c \Phi (t) \gg m_\phi$ and $p_\ast^2 > \delta m_{\text{th};h}^2$, hold
at $\Gamma_\text{inst} \sim H$, one can estimate the ``reheating temperature'' as follows:
\begin{align}
	T_\text{inst} \sim \prn{ \frac{90}{\pi^2 g_\ast} }^\frac{1}{4} \sqrt{\Gamma_\text{inst} \Mpl}
	\simeq
	4 \times 10^{13} \, \text{GeV} \  \com{\frac{c}{0.1}} \com{\frac{0.02}{\alpha_t}}^\frac{1}{4} \com{ \frac{100}{g_\ast} }^\frac{1}{4}
	\com{\frac{m_\phi}{1.5 \times 10^{13}\, \text{GeV}}}^\frac{1}{2},
\end{align}
where $g_\ast$ is the relativistic degrees of freedom produced via this process.
The amplitude of inflaton at that time is estimated as
$\Phi_\text{inst} \sim 10^{15} \,\text{GeV}\ [0.02 / \alpha_t]^{1/2} [c/ 0.1]^2$.
Both conditions, $\pi \alpha_t c \Phi \gg m_\phi$ and $p_\ast^2 > \delta m_{\text{th};h}^2$,
are satisfied for $c \gtrsim 0.1$ at that time.
Moreover, for a large enough coupling $c \gg 0.1$,
the resonance may be terminated by the back-reaction,
and the inflaton itself might participate in thermal plasma.
If this is really the case, the coherence of inflaton is lost, and the resonance becomes inefficient.

In this paper, we do not concretely estimate the coupling $c$ above which
the inflaton participates in the thermal plasma, for the following reasons.
First of all, the quartic interaction yields the Coleman-Weinberg potential~\cite{Coleman:1973jx}:
\begin{align}
	V_\text{CW} = \frac{c^4}{64 \pi^2} \phi^4 \ln \frac{c^2 \phi^2}{m_\phi^2}.
\end{align}
If we stick to the quadratic potential~\eqref{eq:L_inf},
the coupling $c$ is bounded from above $c \lesssim 10^{-3}$~\cite{Lebedev:2012sy}.
In this case, the instant preheating is not efficient.
In addition, for $c \gtrsim 0.1$,\footnote{
	The inflaton-Higgs coupling can affect the RGE running of the Higgs quartic coupling 
	at the energy scale higher than $m_\phi$.
	It may lead to the absolute stability of the Higgs potential for $c\gtrsim 1$, 
	which is out of our interest. 
	Note that the mechanism of Refs.~\cite{Lebedev:2012zw,EliasMiro:2012ay}
	to make the Higgs potential stable does not work in the present model, 
	since the $Z_2$ symmetry $\phi \to -\phi$ is not broken at the present vacuum. 
}
a flattening mechanism of the inflaton potential may be required;
examples are the non-minimal coupling~\cite{Bezrukov:2007ep} or the modified kinetic terms~\cite{Nakayama:2010kt,Kallosh:2013hoa,Kallosh:2013yoa}.
In these cases, the inflaton potential is more complicated, and in particular,
it is dominated by the $\phi^4$ term below a threshold field value of $\phi$
that depends on model parameters.
Since the energy density of inflaton behaves as radiation below this scale,
the resonance does not take place once $p_\ast^2 \sim \delta m_{\text{th};h}^2$ is saturated.
Eventually, the whole system, including inflaton, might be thermalized through thermal dissipations~\cite{Mukaida:2014kpa}.
We postpone this issue to avoid model dependent discussions.

%%%%%%%%%%%%%%%%%%%%%%%%%%%%%%%%
%%%%%%%%%%%% Annihilation %%%%%%%%%%%%
%%%%%%%%%%%%%%%%%%%%%%%%%%%%%%%%
\subsection{Annihilation}
\label{sec:annihilation}

%%%%%%%%%%%%%%%%
\begin{figure}[t]
\begin{minipage}{0.5\hsize}
\begin{center}
\includegraphics[scale = 0.8]{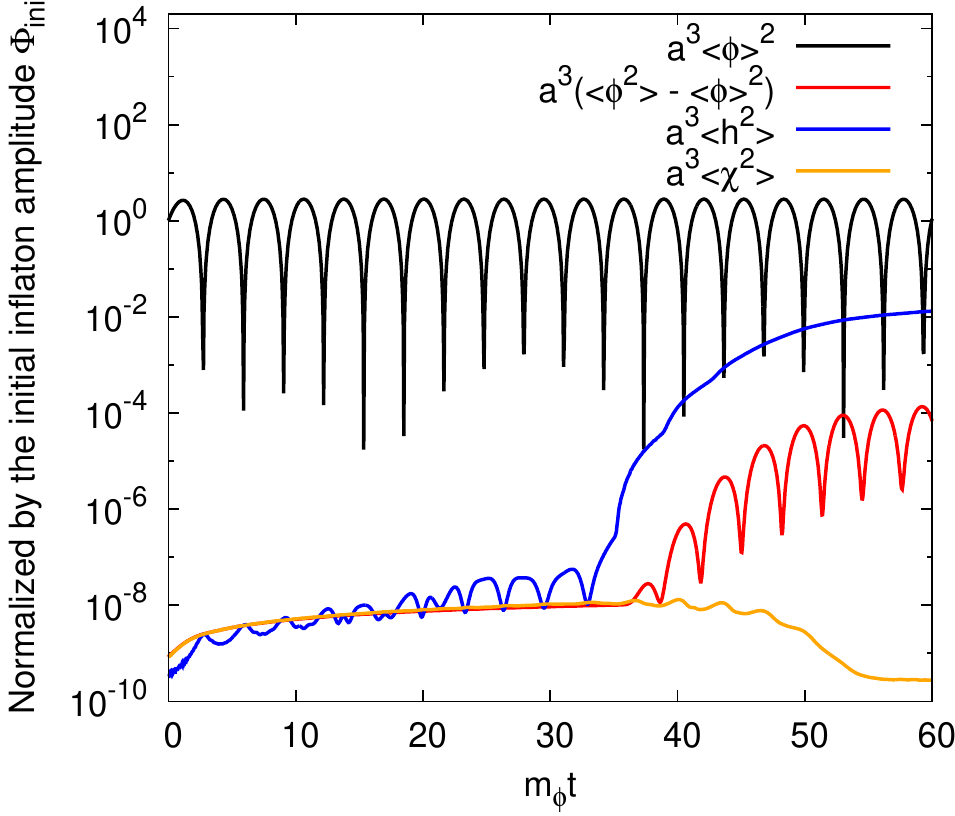}
\end{center}
\end{minipage}
\begin{minipage}{0.5\hsize}
\begin{center}
\includegraphics[scale = 0.8]{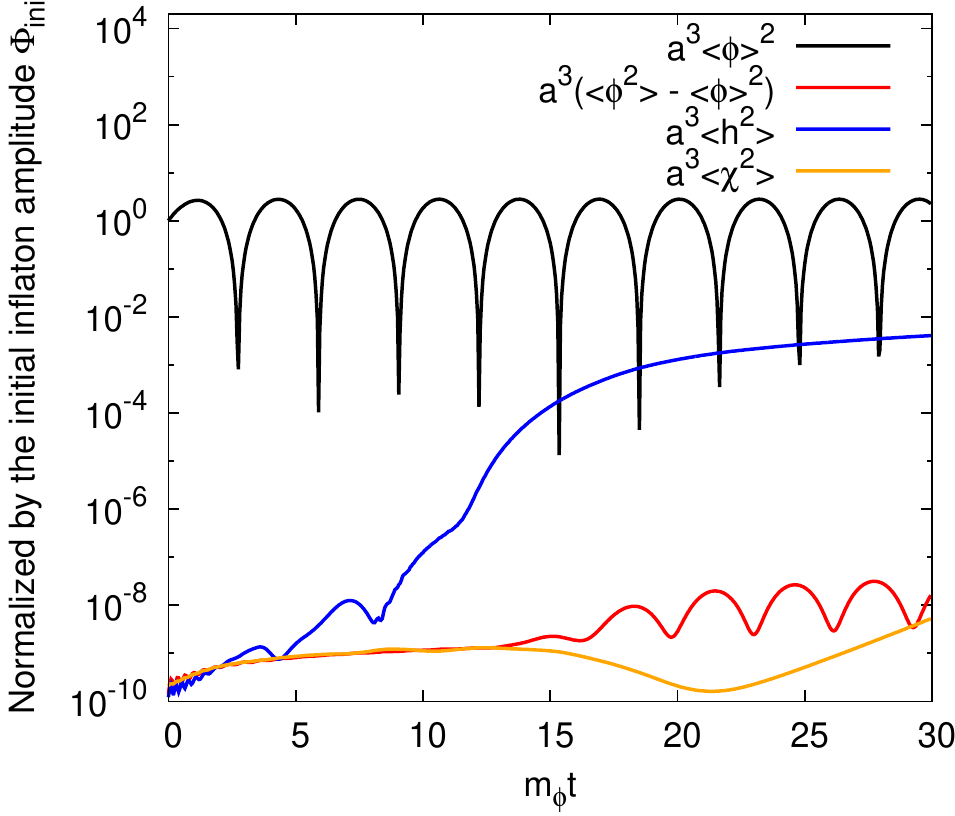}
\end{center}
\end{minipage}
\caption { \small
	Numerical calculation of the time evolution of the inflaton expectation value (black), 
	the inflaton dispersion (\textcolor{red}{red}), the Higgs dispersion (\textcolor{blue}{blue})
	and the $\chi$ dispersion (\textcolor{orange}{orange}). We plot the quartic coupling case for the 
	left panel and the curvature coupling case for the right panel, respectively. We take the parameters
	as $N = 128^3$, $\dd t = 10^{-3}/m_{\phi}$, $m_{\phi} = 1.5\times 10^{13}\,\text{GeV}$,
	$\Phi_{\text{ini}} = \sqrt{2}\,\Mpl$ and $g_{h\chi} = g_{\chi\chi} = 0.5$.
	Left panel: $c = 2\times 10^{-4}$ and $L = 10/m_{\phi}$.
	Right panel: $\xi = 20$ and $L = 20/m_{\phi}$.
	The annihilation processes cannot save the electroweak vacuum.
}
\label{fig:chi}
\end{figure}
%%%%%%%%%%%%%%%%

In the broad/tachyonic resonance, the number density of Higgs grows exponentially.
If the number density is large enough, Higgs can annihilate into top quarks and electroweak gauge bosons.
In particular, Higgs may rapidly excite the gauge bosons to exponentially large number densities~\cite{Felder:2000hr}.
If the number densities of these particles become comparable to that of Higgs before it rolls down to the true vacuum,
they might stabilize Higgs since the gauge coupling is larger than the Higgs quartic coupling.

In order to see whether the annihilation process can stabilize 
the electroweak vacuum during the preheating stage, 
we consider the following simplified Lagrangian:
\begin{align}
	{\cal L} &= {\cal L}_\text{inf} (\phi) + {\cal L}_\text{Higgs} (h) + {\cal L}_{\chi}(\chi)
	+ {\cal L}_\text{int} (\phi, h) + {\cal L}_\text{ann} (h, \chi),
	\label{eq:lag_ann}
\end{align}
where ${\cal L}_\text{inf} (\phi)$, ${\cal L}_\text{Higgs} (h)$ and ${\cal L}_\text{int} (\phi, h)$ are the same 
as those given in Eqs.~\eqref{eq:L_inf}, \eqref{eq:L_higgs} and \eqref{eq:stab}, and
\begin{align}
	{\cal L}_{\chi} (\chi) &=
		\frac{1}{2} \der_\mu \chi \der^\mu \chi - \frac{1}{4}g_{\chi\chi}^2\chi^4, \\[.5em]
	{\cal L}_\text{ann} (h, \chi) &=
		- \ \cfrac{1}{2}\ g_{h\chi}^2 h^2 \chi^2.
\end{align}
Here the light field $\chi$ schematically represents the SM gauge bosons, and we model
the gauge interactions as the quartic interactions.

We have solved the classical equations of motion derived from the Lagrangian~\eqref{eq:lag_ann} numerically.
We take $N = 128^3$, $\dd t = 10^{-3}/m_{\phi}$, 
$m_{\phi} = 1.5\times 10^{13}\,\text{GeV}$, $\Phi_{\text{ini}} = \sqrt{2}\,\Mpl$,
$a_\text{ini} = 1$, $\dot{\Phi}_{\text{ini}} = 0$
and $g_{h\chi} = g_{\chi\chi} = 0.5$.
We summarize some details of our numerical calculation below:
\begin{itemize}

\item 
We introduce gaussian initial fluctuations in the inflaton, Higgs and $\chi$ fields
which arise from the quantum fluctuations.

\item
We renormalize masses of inflaton, Higgs and $\chi$ originating from
the initial quantum fluctuations.
See App.~\ref{app:ren} for more details on this procedure.

\item
We have used the redundancy of the equations of motion for the metric 
as a check of our numerical calculation. 
We have verified that our numerical calculation satisfies
the redundancy at least at $\mathcal{O}(10^{-3})$ precision.

\item We add a sextic term to the Higgs potential to stabilize it in our calculation.
We take the coefficient such that the Higgs field value at the true minimum is $\tilde{\lambda}h_{\rm min}^2 = 5\times 10^{-8}\Mpl^2$.

\item In the case of the curvature coupling,
we solve the equations of motion in the Einstein frame.
We have taken only up to first-order terms 
in $\xi h^2/\Mpl^2$ and $\xi^2 h^2/\Mpl^2$. 
This is because $\xi h^2/\Mpl^2, \xi^2 h^2/\Mpl^2 \ll 1$
always holds in our numerical calculation.

\end{itemize}

In Fig.~\ref{fig:chi}, we show the time evolution of the inflaton vacuum expectation 
value squared $\langle \phi \rangle^2$ (black), 
the inflaton dispersion $\langle \phi^2 \rangle - \langle \phi \rangle^2$ (\textcolor{red}{red}), 
the Higgs dispersion $\langle h^2 \rangle$ (\textcolor{blue}{blue})
and the $\chi$ dispersion $\langle \chi^2 \rangle$ (\textcolor{orange}{orange}). 
They are multiplied by the scale factor to the third.
The left panel is the result of the quartic coupling case with
$c = 2\times 10^{-4}$ and $L = 10/m_{\phi}$,
and the right panel is that of the curvature coupling case with
$\xi = 20$ and $L = 20/m_{\phi}$.
As we can see from Fig.~\ref{fig:chi}, Higgs rolls down to the true vacuum 
well before $\chi$ is sufficiently produced.
In fact, the dynamics of Higgs is almost the same as those in Figs.~\ref{fig:qtc} and \ref{fig:crv}.
Thus, the annihilation process cannot stabilize the electroweak vacuum 
in our simplified setup.
The results do not change even if we take the couplings of $\chi$ larger, 
say $g_{h\chi} = g_{\chi\chi} = 1$.

%%%%%%%%%%%%%%%%
\begin{figure}[t]
\begin{minipage}{0.5\hsize}
\begin{center}
\includegraphics[scale = 0.8]{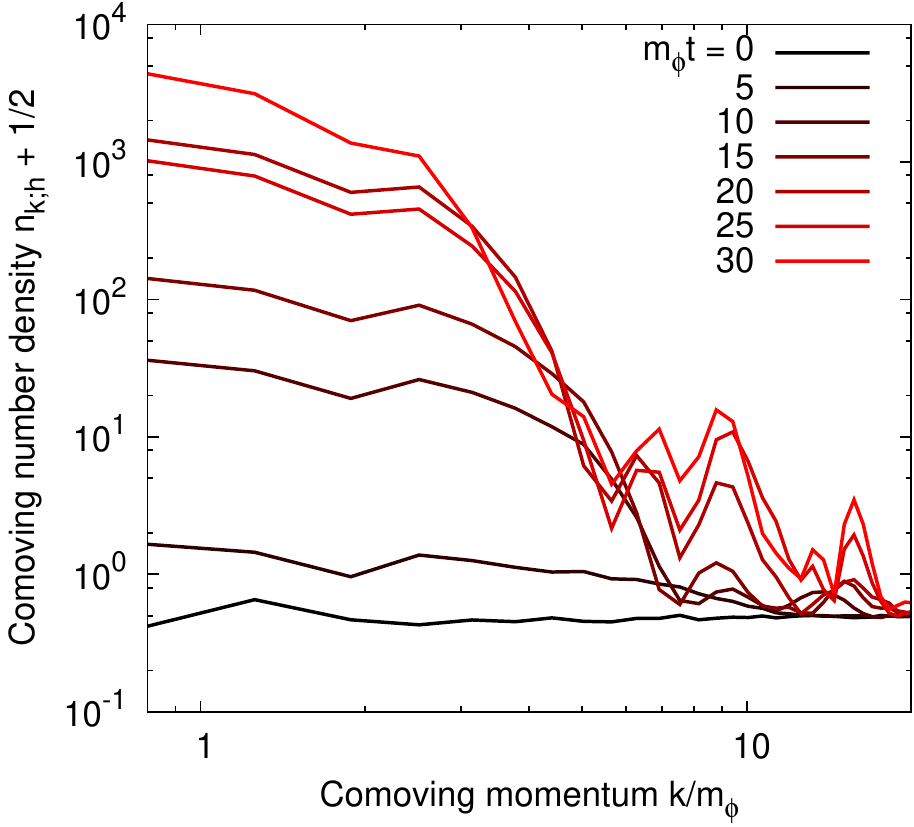}
\end{center}
\end{minipage}
\begin{minipage}{0.5\hsize}
\begin{center}
\includegraphics[scale = 0.8]{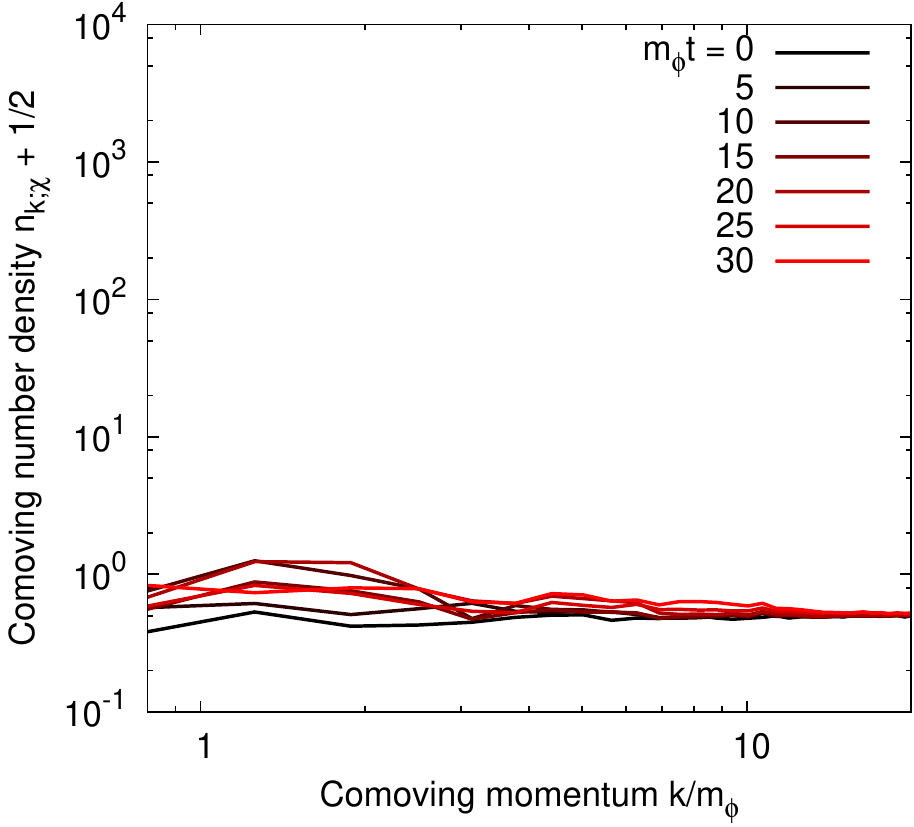}
\end{center}
\end{minipage}
\caption { \small
	We plot the spectra of Higgs and $\chi$
	for the quartic coupling case.
	Left panel: the spectrum of Higgs.
	Right panel: the spectrum of $\chi$.
	Higgs is resonantly produced, 
	while $\chi$ is not efficiently produced.
}
\label{fig:chi_spe}
\end{figure}
%%%%%%%%%%%%%%%%

Now let us take a closer look at Fig.~\ref{fig:chi}.
For the curvature coupling case, Higgs rolls down to the true vacuum 
at the first few oscillation of inflaton,
regardless of the existence of the annihilation channel.
It seems that this is because the time scale of the electroweak vacuum decay 
is too short for the annihilation to be effective.
We believe that the same argument also holds for the realistic model where
Higgs couples to the electroweak gauge bosons, since the annihilation rate is
not much different from that of our simplified model.

For the quartic coupling case, Higgs remains in the electroweak vacuum for the first
$\mathcal{O}(1\mathchar`-10)$ times of oscillation of inflaton.
Even in such a case, however, $\chi$ is not efficiently produced via the annihilation process.
To see this, we plot the spectrum of Higgs and $\chi$ for the quartic coupling case in Fig.~\ref{fig:chi_spe}.
We can see that Higgs is resonantly produced, while $\chi$ is not efficiently produced.
This behavior may be understood as follows.
If Higgs is resonantly produced, it induces an effective mass to $\chi$ as a finite density correction:
\begin{align}
	m_{\text{eff};\chi}^2(t)
	= g_{h\chi}^2 \int_{{\bm k}/a(t)} \frac{1}{2}G_{F;h}(t,t; {\bm k}) 
	\simeq g_{h\chi}^2 \frac{n_{h}(t)}{\omega_{k_*;h}(t)}.
\end{align}
The phase space of the produced $\chi$ is 
kinematically restricted by the effective mass $m_{\text{eff};\chi}^2$,
and hence it may suppress the annihilation rate.\footnote{
	As an extreme case, the annihilation process is 
	kinematically forbidden if $m_{\text{eff};\chi} \gg p_*$.
}
Provided that this description is correct, 
the dynamics should be similar for the realistic case
where Higgs couples to the gauge bosons.
The resonant production of Higgs induces effective masses to the gauge bosons,
and hence we expect that the annihilation of Higgs 
into the gauge bosons is suppressed in the same way as our simplified model.
Thus, we believe that
the annihilation process cannot stabilize the electroweak vacuum 
during the preheating stage in the realistic case either.
More rigorous analysis requires classical lattice simulations 
including SM gauge bosons~\cite{Figueroa:2015rqa,Enqvist:2015sua} and fermions, which is beyond the scope of this paper.

Note that, even if we properly take into account the degrees of freedom 
of the gauge boson and Higgs, the situation is not expected to change.
The reason is the following. In order for the gauge boson to stabilize Higgs,
its dispersion must satisfy
\begin{align}
	g^2 \sum \left\langle A^2 \right\rangle \sim \lvert \lambda \rvert \sum \left\langle h^2 \right\rangle,
	\label{eq:stb_cond_anni}
\end{align}
where the sums are for the six degrees of freedom for the gauge boson $A$ and the four degree of freedom for Higgs $h$, 
respectively.
Expecting that $\left \langle A^2 \right\rangle \simeq \left \langle \chi^2 \right\rangle$,
we can see from Figs.~\ref{fig:chi} and~\ref{fig:chi_spe} that
the produced amount of the gauge boson is far below Eq.~\eqref{eq:stb_cond_anni}.

It is noticeable that it depends on the numerical value of the Higgs quartic coupling 
whether the annihilation can stabilize the electroweak vacuum or not.
In the present case with the center value of top quark mass, 
the Higgs quartic coupling is rather large, $\tilde{\lambda} \sim 0.01$,
and hence the annihilation cannot catch up the vacuum decay. 
However, it can stabilize the electroweak vacuum if $\tilde{\lambda}$ is small enough. 
It may be worth studying further on this respect.

%%%%%%%%%%%%%%%%%%%%%%%%%%%%%%%%
%%%%%%%%%%% After preheating %%%%%%%%%%%
%%%%%%%%%%%%%%%%%%%%%%%%%%%%%%%%
\section{After Preheating}
\label{sec:aft_prh}

So far we have focused on the stability of Higgs during the preheating stage. 
We have obtained the robust bound on the Higgs-inflaton coupling
above which Higgs rolls down to its true vacuum.
In this section, we consider the dynamics of Higgs after the preheating stage
in the case where the electroweak vacuum survives the preheating stage.
We study whether or not the dynamics after the preheating forces the electroweak vacuum to decay.
For notational simplicity, we denote the effective mass of Higgs induced by 
the Higgs-inflaton/-curvature coupling as $m_{H;h}^2$.
It is given by $m_{H;h}^2 = c^2\Phi^2(t)/2$ for the Higgs-inflaton coupling
and $m_{H;h}^2 = \xi m_{\phi}^2 \Phi^2(t)/2\Mpl^2$ for the Higgs-curvature coupling
[See Eqs.~\eqref{eq:disp_quart} and \eqref{eq:disp_crv}].

First we summarize basic properties of Higgs fluctuations just after the preheating stage.
At $t = t_\text{end}$,
$m_{H;h}$ is comparable to the inflaton mass, $m_{H;h} (t_\text{end}) \simeq m_\phi$,
The physical momentum scale of Higgs is estimated as 
$p_\ast (t_\text{end}) \simeq m_\phi \simeq m_{H;h} (t_\text{end})$.
Since we assume that
the electroweak vacuum survives the preheating stage, 
the tachyonic effective mass term of Higgs induced by its self interaction satisfies
$\lvert\delta m_{\text{self};h}(t_{\text{end}})\rvert < p_\ast (t_\text{end}) \simeq m_{H;h} (t_\text{end})$.
Therefore, all wave length modes are stable against 
the tachyonic mass term just after the preheating.\footnote{
	See the $q < 1$ and $0 < A_{k} < 1$ region of the stability/instability chart of the Mathieu equation
	in Ref.~\cite{Tsujikawa:1999jh}.
} 
As one can guess from Figs.~\ref{fig:spe_qtc} and \ref{fig:spe_crv},
the comoving number density in a momentum space at $t = t_\text{end}$ may be approximated as
\begin{align}
	n_{k;h} (t_\text{end}) = 
	\begin{cases}
	f &\text{for}~~~H \ll  k/ a_\text{end} \lesssim m_\phi \\[.5em]
	k^{-n} &\text{for}~~~m_\phi \lesssim  k/ a_\text{end}
	\end{cases},
	\label{eq:dist_aft_prh}
\end{align}
with $n \geqslant 4$.
The stability condition, $m_\phi^2 > |\delta m_{\text{self};h}^2 (t_\text{end})|$, puts a rough upper bound
$f \lesssim 10^{3} \times  (10^{-2} / \tilde \lambda)$.
A characteristic coupling of this system is roughly, $\alpha, \alpha_t \sim 0.02$--$0.1$,
where $\alpha$ is the fine structure constant of the SM gauge group.
Let us parametrize the typical value of Higgs distribution as $f = \alpha^{-m}$.
The Higgs distribution mostly lies in the ``\textit{over-occupied}'' regime, $0 < m <1$; or
``\textit{extremely over-occupied}'' regime, $1 < m$~\cite{Kurkela:2011ti},  
at the boundary of the inequalities~\eqref{eq:claim_qtc} and \eqref{eq:claim_crv}.

A complete analysis of this system after the preheating stage is beyond the scope of paper.
Instead, we discuss its possible processes, and point out important ingredients which could change the dynamics qualitatively.

\subsection{Cosmic expansion}
\label{sec:cosmic_exp}

First, let us discuss the effect of cosmic expansion.
As an illustration, let us 
consider only the effective masses induced by the Higgs-inflaton coupling and the Higgs self interaction.
The mass term $m_{H;h}^2$ decreases as $m_{H;h}^2 \propto a^{-3}$
since the inflaton harmonically oscillates with time.
On the other hand, the effective mass of Higgs induced by the its self coupling follows
$\delta m_{\text{self};h}^{2} \propto a^{-2}$. This is because the modes with 
the momentum $p \sim p_{*}$ or $p_{*}^{(\text{tac})}$ dominate the Higgs dispersion,
and hence we can treat them as relativistic particles.

Right after the end of the preheating, the mass term $m_{H;h}^2$ stabilizes the electroweak vacuum
against the tachyonic mass term $\delta m_{\text{self};h}^2$ 
generated by the Higgs self interaction.
However, since $m_{H;h}^2$ decreases faster than $\delta m_{\text{self};h}^2$,
the long wave length mode of Higgs may be destabilized eventually.
We now estimate the time scale $t_\text{eq}$
when the inflaton induced mass becomes comparable to the tachyonic mass.
For the quartic stabilization, we get
\begin{align}
	m_\phi t_\text{eq} \sim 6 \times 10^3 \ 
	\com{ \frac{c}{10^{-4}} }^\frac{7}{4} \exp \com{ -7.8 \times \prn{ \frac{c}{10^{-4}} - 1 } }.
	\label{eq:teq_quart}
\end{align}
Here we focus on the coupling $c$ dependence, and take 
$\mu_\text{qtc} = 0.1$ and $m_\phi = 1.5 \times 10^{13}$ GeV.
For the curvature stabilization, the time scale is estimated as
\begin{align}
	m_\phi t_\text{eq} \sim 2 \times 10^3\  \com{ \frac{\xi}{2} }^\frac{3}{4}
	\com{ \frac{\Phi_\text{ini}}{\sqrt{2}\Mpl} }^\frac{1}{2}
	 \exp \com{ - 6 \times  \prn{ \sqrt{\frac{\xi}{2}} \frac{n_\text{eff}\,\mu_\text{crv}}{2} \frac{\Phi_\text{ini}}{\sqrt{2}\Mpl} - 1}}.
	 \label{eq:teq_curv}
\end{align}
Here we take $m_\phi = 1.5 \times 10^{13}$ GeV.

If the Higgs self coupling $\lambda$ is still negative at $t = t_\text{eq}$, 
Higgs eventually rolls down to the true vacuum.
For the electroweak vacuum not to decay after the preheating, 
$\langle h^2 \rangle$ should be smaller than $h_{\rm max}^2$
at $t = t_{\rm eq}$.
Thus, we derive the upper bounds on the couplings as
\begin{align}
	c 
	&\lesssim 
	3 \times 10^{-5} \  \com{\frac{0.1}{\mu_{\text{qtc}}}}\ \com{\frac{m_\phi}{1.5 \times 10^{13} \,\text{GeV}}}
	\com{1 + 0.4 \ln \prn{ \frac{h_{\text{max}}/10^{10}\,\text{GeV}}{m_{\phi}/1.5\times10^{13}\,\text{GeV}} }} ,
	\label{eq:claim_qtc_aft}
\end{align}
for the quartic stabilization, and
\begin{align}
		\xi &\lesssim 
		7 \times 10^{-1} \com{\frac{2}{n_{\text{eff}}\,\mu_{\text{crv}}}}^2 \com{\frac{\sqrt{2}\,\Mpl}{\Phi_\text{ini}}}^2
		\com{1 + 0.3 \ln \prn{ \frac{h_{\text{max}}/10^{10}\,\text{GeV}}{m_{\phi}/1.5\times10^{13}\,\text{GeV}} }}^2, 
	\label{eq:claim_crv_aft}
\end{align}
for the curvature stabilization.\footnote{
It should be regarded as an illustration since 
the condition for the resonance to occur is only marginally
satisfied at these values.
}
Interestingly, 
almost all the parameters required to stabilize the electroweak vacuum 
during inflation results in the catastrophe,
unless there exist other contributions to the Higgs effective mass term.
Thus, the inclusion of Higgs-radiation coupling is crucial.

\subsection{Turbulence and thermalization}
\label{sec:turb}

Next, let us discuss how the Higgs-radiation coupling could change these bounds qualitatively.
In this section, we neglect radiation generated in the process of complete reheating,
assuming that the reheating temperature is low enough.
See also the discussion in the next subsection.

Higgs
couples with electroweak gauge bosons and top quark
via gauge and top Yukawa couplings. 
The initial Higgs distribution illustrated in Eq.~\eqref{eq:dist_aft_prh}
is eventually thermalized while producing SM particles, and
its thermalization time scale is characterized by these couplings.
Once it is thermalized, the lifetime of our vacuum can be estimated 
by means of the thermal bounce~\cite{Espinosa:1995se,Ellis:2009tp}.
In particular, it was shown in Ref.~\cite{Rose:2015lna} that the lifetime of the electroweak vacuum is long enough 
if we adopt the central value of the top quark mass.
Therefore, we expect that the electroweak vacuum is stable 
if thermalization time scale is much shorter than $t_\text{eq}$.

Now we roughly estimate the thermalization time scale.
As shown in Refs.~\cite{Micha:2002ey,Micha:2004bv,Berges:2008wm},
an initially over-occupied system, like Eq.~\eqref{eq:dist_aft_prh}, enters the turbulent regime at first,
cascades self-similarly, and eventually attains thermal distribution.
In addition, the IR cascade may develop the long wave length mode as pointed out in Ref.~\cite{Berges:2008wm,Berges:2012us}, 
and might boost the vacuum decay.
Here, however, as an illustration,
we simply compare the typical time scale of elastic scatterings with the Hubble parameter.
In fact, for a mildly over-occupied system, 
its thermalization may be dominated by the elastic scatterings~\cite{Kurkela:2011ti,Micha:2004bv}.
This is the case of $c \lesssim 10^{-4}$ for the quartic coupling
and $\xi \lesssim 1$ for the curvature coupling, though it strongly depends on which interaction dominates
the thermalization. 
The thermalization time of elastic scatterings may be evaluated as $\alpha^2 T^\text{(w.b.)} (t_\text{th}) \sim H (t_\text{th})$
with $T^\text{(w.b.)} (t)$ being an would-be temperature when the system at that time $t$ would be thermalized.
We find that the thermalization time scale, $t_\text{th}$, is much longer than or at most comparable to $t_\text{eq}$
for the parameter region of our interest.
Further studies on this case will be presented elsewhere.

\subsection{Complete reheating}
\label{sec:reheating}

Finally, we study the effective mass of Higgs from radiation generated during the process of
the complete reheating.
Since the Higgs-inflaton/-curvature coupling alone cannot lead to a complete decay of inflaton,
an additional interaction of inflaton which completes the reheating is required.
We discuss how it affects the dynamics of Higgs both during and after the preheating in the following.

To discuss the complete reheating process in detail~\cite{Harigaya:2013vwa,Mukaida:2015ria},
we have to specify the interaction between inflaton and radiation,
which strongly depends on inflationary models.
In the following discussion, we focus on the case where inflaton reheats the Universe
via Planck-suppressed operators; the decay rate of inflaton is given by
\begin{align}
	\Gamma_\phi =  \frac{\tilde \Gamma_\phi m_\phi^3}{\Mpl^2}.
\end{align}
Here $\tilde \Gamma_\phi$ is a constant that is smaller than unity $\tilde \Gamma_\phi \ll 1$.
For a dimension-five Planck-suppressed decay of inflaton, 
we may expect $\tilde \Gamma_\phi \sim \mathcal O (0.1)$.
At the end of the next sub-section, we briefly comment on the case with slightly larger couplings between
inflaton and radiation.

The ``thermal mass'' contribution to the Higgs effective potential may be parametrized as
\begin{align}
	\frac{1}{2}\delta m_\text{th;h}^2 h^2 \theta \prn{ h_\text{th} - h},
\end{align}
where $ \theta$ is the Heaviside step function, $\delta m_{\text{th};h}$ is the ``thermal mass''
from radiation generated during the course of the complete reheating,
and $h_\text{th}$ is a typical threshold field value above which the electroweak gauge bosons and top quark may not
be produced efficiently owing to its large effective mass proportional to the field value of Higgs.
Both $\delta m_{\text{th};h}$ and $h_\text{th}$ depend on time.
See Appendix~\ref{app:thrm_aft_inf} for details, and the concrete forms of $\delta m_{\text{th};h}$ and $h_\text{th}$ 
for each regime, given in Eqs.~\eqref{eq:m_th_hard} and \eqref{eq:m_th_soft}.
We expect that the electroweak vacuum is stabilized if both
$\delta m_{\text{th};h}^2 (t_\text{eq}) > \lvert \delta m_{\text{self};h}^2 (t_\text{eq}) \rvert$ 
and $h_\text{th} (t_\text{eq}) > \sqrt{\langle h^2 (t_\text{eq}) \rangle}$
are fulfilled.\footnote{
	This requirement is somewhat conservative,
	for the ``thermal mass'' from hard primaries [Eq.~\eqref{eq:m_th_hard}]
	may reduce the Higgs dispersion at the early stage of preheating
	in the case of $|\delta m_{\text{self};h}| < p_\ast$.
}

\subsubsection*{During preheating}

Here we discuss whether or not the dynamics of the complete reheating would change
the upper bound given in Eqs.~\eqref{eq:claim_qtc} and \eqref{eq:claim_crv}.

First of all, let us estimate a relevant time scale in the following discussion.
For the quartic stabilization, the longest time scale is governed by
Eq.~\eqref{eq:stab}, which characterizes the end of resonant amplification by the cosmic expansion:
\begin{align}
	m_\phi t_\text{end} \simeq 40 \times \Bigg(  \frac{c}{10^{-4}} \Bigg) \prn{ \frac{10^{13}\, \text{GeV}}{m_\phi} }.
\end{align}
For the curvature stabilization, the dynamics is almost determined by the first few oscillations.
Hence, we have to deal with the time scale of ${\cal O} (10^{0 - 2})$ times oscillations of inflaton.
Since this time scale is rather short, we cannot rely on the instantaneous thermalization approximation,
frequently assumed in literature~\cite{Espinosa:2015qea,Giudice:2000ex}.
To see this, it is instructive to roughly estimate the two-to-two scattering rate
because, at least, this interaction should be faster than the cosmic expansion
to attain thermal equilibrium.
A naive estimation may give the rate of 
$(\alpha^2/m_s^2) \times n_\text{rad} \sim \alpha m_\phi$, where $\alpha$ is the fine structure constant of SM gauge group
and the screening mass is given by $m_s^2 \sim \alpha n_\text{rad}/m_\phi$.
One can see that even the two-to-two scattering does not take place for $m_\phi t \lesssim \alpha^{-1} \sim {\cal O} (10)$,
which is comparable to the time scale of our interest.
The above estimation is too naive, so it should be understood as an illustration.
See Appendix~\ref{app:thrm_aft_inf} for details.

Let us estimate the effects of radiation produced during the process of complete reheating.
Radiation induces the effective potential 
$\sim \delta m_{\text{th};h}^2 h^2 \theta (h_\text{th} - h)$
with $\delta m_{\text{th};h}^2$ and $h_\text{th}$ depending on time $t$
[See Eqs.~\eqref{eq:m_th_hard} and \eqref{eq:m_th_soft}].
This effective mass is always smaller than the inflaton mass,
and thus the condition for the efficient Higgs production 
given in Eq.~\eqref{eq:np_cond} [Eq.~\eqref{eq:tac_cond}] holds: $p_\ast > m_\phi \gg \delta m_{\text{th};h}$.
Also, the condition given in Eq.~\eqref{eq:ema_cond}  [Eq.~\eqref{eq:ema_cond_2}] holds for $p_\ast > m_\phi$
because of $|\delta m_{\text{self};h}| > p_\ast > m_\phi \gtrsim \tilde \lambda^{1/2} h_\text{th}$.
Therefore, in the case of  Planck-suppressed decay of inflaton,
we expect that 
the process of complete reheating is not likely to
change the upper bound given in Eqs.~\eqref{eq:claim_qtc} and \eqref{eq:claim_crv}.
However, note that we have taken a sufficiently small $\tilde \Gamma_\phi$
so that resonant particle production of SM particles other than Higgs does not occur.
An enhanced $\tilde \Gamma_\phi$ might affect the obtained bound.
See also discussion in Sec.~\ref{sec:conclusion}.

\subsubsection*{After preheating}

The dynamics of Higgs after the preheating strongly depends on the reheating dynamics.
This is because radiation produced via the complete reheating
generates additional Higgs effective mass term, which follows $\delta m_{\text{th};h} \propto a^{-3/8}$
before the complete decay of inflaton.
Since it decreases much slower than the tachyonic mass term,
this term eventually takes over the dominant contribution of the effective mass.

First, let us discuss a case with $T_\text{RH} \simeq 10^{10}$ GeV,
which is a typical example of dim.\ 5 Planck-suppressed decay.
For the quartic stabilization with 
$c \simeq 10^{-4}$, a typical time scale given in Eq.~\eqref{eq:teq_quart}
resides in $m_\phi t_\text{eq} \sim 6 \times 10^3 \subset [\tilde t_\text{max}, \tilde t_\text{RH}]$
for $T_\text{RH} \simeq 10^{10}$ GeV [See Eq.~\eqref{eq:m_th_soft} and definitions below it].
The thermal mass of Higgs at that time is given by
\begin{align}
	\delta m_{\text{th};h} (t_\text{eq}) \sim 
	10^{12} \, \text{GeV} \ \com{\frac{\alpha}{0.1}}^\frac{1}{2} \com{\frac{\tilde \Gamma_\phi}{0.2}}^\frac{1}{4}
	\com{\frac{6 \times 10^3}{m_\phi t_\text{eq}}}^\frac{1}{4},
\end{align}
with $k_\text{max} \sim 10^{12}$ GeV.
The effective mass term of Higgs induced from its self interaction at that time is
\begin{align}
	\abs{\delta m_{\text{self};h} (t_\text{eq})} \sim 7 \times 10^{10} \, \text{GeV} \
	\com{\frac{10^{-4}}{c}}^\frac{3}{4} \exp \com{ 7.8 \times \prn{ \frac{c}{10^{-4}} -1} }.
\end{align}
For a smaller coupling of $c$,
the equality time, $t_\text{eq}$, becomes longer,
and it is more likely to be stabilized by the thermal contribution,
for the thermal mass decreases slower than the effective mass of Higgs generated from its self interaction.
Therefore, we expect that
the reheating temperature of $T_\text{RH} \simeq 10^{10}$ GeV may save the electroweak vacuum
in the case with a quartic coupling of $c \lesssim 10^{-4}$,
because the conditions, $\delta m_{\text{th};h} > |\delta m_{\text{self};h}|$ and 
$h_\text{th}=k_\text{max}/g > |\delta m_{\text{self};h}| / \tilde \lambda^{1/2}$, are satisfied.
Also, for the curvature stabilization with $\xi \simeq 2$,
the typical time scale shown in Eq.~\eqref{eq:teq_curv}
resides in $m_\phi t_\text{eq} \sim 1.6 \times 10^3 \subset [\tilde t_\text{el}, \tilde t_\text{soft}]$
for $T_\text{RH} \simeq 10^{10}$ GeV [See Eq.~\eqref{eq:m_th_soft} and definitions below it].
The ``thermal mass'' of Higgs may be given by
\begin{align}
	\delta m_{\text{th};h} (t_\text{eq}) \sim 
	10^{12} \, \text{GeV} \ \com{\cfrac{\alpha}{0.1}}^\frac{3}{4} \com{\cfrac{\tilde \Gamma_\phi}{0.2}}^\frac{3}{8}
	\com{\cfrac{1.6 \times 10^3}{m_\phi t_\text{eq}}}^\frac{1}{4},
\end{align}
with $k_\text{max} \sim 6 \times 10^{11}$ GeV.
The effective mass term of Higgs induced from its self interaction is estimated as
\begin{align}	
	\abs{\delta m_{\text{self};h} (t_\text{eq})} \sim 2 \times 10^{10} \, \text{GeV} \
	\com{ \frac{\xi}{2} }^{-\frac{1}{4}} \com{ \frac{\sqrt{2}\Mpl}{\Phi_\text{ini}} }^\frac{1}{2}
	\exp \com{6 \times \prn{ \sqrt{\frac{\xi}{2}} \frac{\Phi_\text{ini}}{\sqrt{2} \Mpl} \frac{n_\text{eff} \mu}{2} -1 }}.
\end{align}
Hence, we expect that 
the reheating temperature of $T_\text{RH} \simeq 10^{10}$ GeV may save the electroweak vacuum
in the case with a curvature coupling of $\xi \lesssim 2$.
Note, however, that we have assumed $\xi \gg 1$ in our estimation,
and hence the numerics, $\xi \simeq 2$, should be understood as an illustration.

Next, let us estimate the lower bound of the reheating temperature
below which the reheating dynamics cannot save the electroweak vacuum,
utilizing the results given in \cite{Harigaya:2013vwa,Mukaida:2015ria}.
In order to estimate the lower bound conservatively,
we demand that the ``thermal mass'' term, $\delta m_{\text{th};h}^2$, is always much smaller than the dispersion of Higgs,
$\langle h^2 \rangle$, throughout the thermal history up to the equality time $t_\text{eq}$.
Moreover, strictly speaking,
not only the thermal potential but the thermal dissipation might relax the Higgs to its enhanced symmetry point.
A complete analysis is beyond the scope of this paper.
Instead, we simply require that the thermal interactions are slow enough $\alpha^2 T t_\text{eq} \ll 1$,
which is essentially the same as $m_\phi t_\text{eq} \ll \tilde t_\text{soft}$.
Imposing these requirements,
for both the quartic/curvature stabilization,
we obtain a rather conservative bound on the reheating temperature below which
the reheating dynamics cannot save the electroweak vacuum:
$T_\text{RH} \lesssim {\cal O} (10^5)$ GeV.
Anyway, we need further investigations to derive more precise lower bound on the reheating temperature.

%%%%%%%%%%%%%%%%%%%%%%%%%%%%%%%%
%%%%%%%%%%%% Conclusion %%%%%%%%%%%%
%%%%%%%%%%%%%%%%%%%%%%%%%%%%%%%%
\section{Conclusions and Discussion}
\label{sec:conclusion}

The current experimental data of the Higgs and top quark masses indicates
that the electroweak vacuum is metastable 
if there is no new physics other than the SM.
From the viewpoint of inflationary cosmology,
one interesting consequence is that high-scale inflation requires some stabilization mechanism of Higgs during inflation. 
A possible candidate of such a mechanism is the Higgs-inflaton/-curvature coupling.
In fact, it induces an effective mass and stabilizes Higgs during inflation.
After inflation,
however, it causes an exponential enhancement of Higgs fluctuations due to the broad/tachyonic resonance,
and hence the electroweak vacuum may eventually decay into the true one during the preheating stage.

In this paper, we have focused on the preheating dynamics of Higgs 
induced by the Higgs-inflaton/-curvature coupling.
We have clarified in what parameter space our electroweak vacuum
decays into the true one via the broad/tachyonic resonance.
We have derived the criterion when Higgs rolls down to the true vacuum, 
and confirmed it by performing the
$3+1$-dimensional classical lattice simulations. To be concrete, 
the electroweak vacuum survives the preheating stage
only if the couplings satisfy the following inequalities:
\begin{align*}
	c &\lesssim  10^{-4} \  \com{\frac{0.1}{\mu_{\text{qtc}}}}\ \com{\frac{m_\phi}{10^{13} \,\text{GeV}}},
\end{align*}
for the quartic coupling case,
and
\begin{align*}
	\xi &\lesssim 10  \com{\frac{2}{n_{\text{eff}}\,\mu_{\text{crv}}}}^2 \com{\frac{\sqrt{2}\,\Mpl}{\Phi_\text{ini}}}^2,
\end{align*}
for the curvature coupling case,
as long as the broad/tachyonic resonance is effective at the onset of the inflaton oscillation. 
See Eqs.~\eqref{eq:claim_qtc} and \eqref{eq:claim_crv}.
These conditions claim that the Hubble expansion should kill the resonance 
before the Higgs self coupling becomes relevant. 
In order to suppress the fluctuation of Higgs during inflation, 
the couplings $c$ and $\xi$ should satisfy
$c \gtrsim {\cal O}(H_\text{inf}/\Phi_{\text{ini}})$ and 
$\xi \gtrsim {\cal O}(0.1)$~\cite{Espinosa:2007qp,Espinosa:2015qea}.
Thus, our results indicate that the Higgs-inflaton/-curvature coupling
should be rather small to stabilize Higgs during both the inflation and the preheating stages.
We have also seen that the Higgs-radiation coupling does not change the situation
as long as the inflaton \textit{perturbatively} reheats the Universe,
\textit{i.e.} no resonant particle production occurs except for Higgs.
This is the main conclusion of this paper.

Here we give some remarks.
First of all, we comment on the dynamics of Higgs after the preheating,
in the case where the electroweak vacuum survives during the preheating stage.
As explained in Sec.~\ref{sec:cosmic_exp},
if one neglects the Higgs-radiation coupling, the cosmic expansion leads to the decay of our electroweak vacuum
for almost all the parameters of our interest [See Eqs.~\eqref{eq:claim_qtc_aft} and \eqref{eq:claim_crv_aft}].
Thus, including the Higgs-radiation coupling is important to discuss the fate of electroweak vacuum after the preheating.
Such an over-occupied system, as illustrated in Eq.~\eqref{eq:dist_aft_prh}, may exhibit turbulence and
cascade towards not only UV but IR, which might have implications on thermalization after the preheating.
As a first step,
we have simply compared the elastic scattering rate with the Hubble parameter
and have seen that the cosmic expansion is faster.
However, in order to estimate the conditions to avoid the catastrophe quantitatively,
we might have to perform numerical simulations including Higgs-radiation coupling.
Moreover, we have mentioned in Sec.~\ref{sec:reheating}
that the coupling between inflaton and radiation which leads to the complete reheating plays the crucial role. 
Importantly, the relevant time scale of Higgs dynamics is rather short,
and the instantaneous thermalization assumption of radiation
might be questionable, in particular for a low reheating temperature.
We have roughly estimated its effect in two cases; typical reheating temperature of $ T_\text{RH} \simeq 10^{10}$ GeV
and low reheating temperature of $ T_\text{RH} \simeq 10^{5}$ GeV.
It was shown that thermal effects might save the electroweak vacuum in the former case
after the preheating, while in the latter case, the cosmic expansion kills almost all the parameters required for
the stability of vacuum during inflation.
However, since it crucially depends on thermalization processes, further studies are required
to obtain quantitative results.

Second, we have assumed that other SM particles than Higgs are produced \textit{perturbatively} via
the decay of inflaton, and neglected their resonant production.
However, the resonance takes place at the early stage of the complete reheating,
for instance,
in the case of $(\phi / \Mpl) \tilde F F$ with a sizable coupling,
which yields $T_\text{RH} \gtrsim {\cal O} (10^{10} )$ GeV.
Their resonant production might affect the stabilization of Higgs during the preheating stage,
although its efficiency strongly depends on 
couplings of inflaton with radiation and inflaton amplitude after the inflation.
We leave thorough studies in this respect as a future work.

Third, we comment on the $h_\text{max}$ dependence of our result.
We have used the value $h_\text{max} = 10^{10}\GEV$ in this paper, 
but we expect that our result does not change much
as long as $h_{\text{max}}$ satisfies $\tilde{\lambda}h_\text{max}^2 \ll \sqrt{q}m_\phi^2$ or $qm_\phi^2$
for the quartic or curvature coupling case, respectively.
If the inequality is inverted, the resonance shuts off due to the positive Higgs self-coupling
even for a quite large value of the resonance parameter. 
Thus, Higgs will be trivially stable against the preheating in such a case.
Precise determination of the top mass will make the situation clearer in the future.

Fourth, we comment on a possible effect of a tail of the Higgs distribution.
In this study, we have estimated the condition where the electroweak vacuum decays on average.
In reality, however, we have $e^{3\mathcal{N}}$ numbers of Hubble patches with 
$\mathcal{N} \simeq 50 \mathchar`- 60$,
and hence we must estimate the condition at which no one in the $e^{3\mathcal{N}}$ Hubble patches
experiences the electroweak vacuum decay.
It is possible that the conditions~\eqref{eq:ema_cond} and~\eqref{eq:ema_cond_2} 
are violated in one Hubble patch even
if they are satisfied on average since the distribution of the Higgs field value has a tail.
Thus, the condition for $c$ and $\xi$ can be even severer once we include this effect.
A detailed study on this respect requires a precise knowledge about the distribution of Higgs, and
we leave it as a future work.

Fifth, we can generalize our study to the following Planck-suppressed interaction:
\begin{align}
	{\cal L}_\text{int} = -\frac{h^2}{6\Mpl^2}\com{\frac{c_{K}}{2}\der_\mu \phi \der^\mu \phi+ c_{V}V\prn{\phi}},
\end{align}
where $V\prn{\phi}$ is the potential of inflaton,
although we have treated only the Higgs-inflaton and -curvature coupling given in Eq.~\eqref{eq:stb_intro} in this paper.
If $c_{K} \simeq c_{V}$, the effective mass term of Higgs induced by this interaction does not oscillate much
and hence the resonant Higgs production is expected to be suppressed, leading to weaker constraint.\footnote{
	Precisely speaking, even if $c_K = c_V$, the energy density of inflaton and the Hubble parameter have
	oscillating parts at the onset of the oscillation~\cite{Ema:2015dka}.
}
If $c_{K} \neq c_{V}$, the effective mass term oscillates with time during the inflaton oscillation regime,
and we obtain similar constrains to the case studied in the main text.
Note that the these couplings generically exist due
to, \textit{e.g.} radiative processes. For example, it is discussed in Ref.~\cite{Gross:2015bea} 
how the Higgs-inflaton quartic coupling emerges from loop effects in various models. 
The Higgs-curvature coupling is also generally induced 
by radiative corrections in the curved space~\cite{Herranen:2014cua}.

Finally, we comment on the shape of the inflaton potential. In this paper, we assumed that 
inflaton oscillates around the origin of the potential, which is typical in high-scale inflation models. 
However, it is possible that inflaton oscillates around some finite vacuum expectation value (VEV). 
The result does not change in the case of the Higgs-curvature coupling:
we can just regard $\phi (t)$ as the displacement from the VEV. 
In the case of quartic coupling $c^2\phi^2h^2$, the results depend on the value of inflaton VEV
around which the inflaton oscillates.
It is expected that the resonant Higgs production effect becomes weaker for larger VEV,
although further detailed investigations will be necessary to derive precise constraints on parameters.

%%%%%%%%%%%%%%%%%%%%%%%%%%%%%%%%
%%%%%%%%%% Acknowledgements %%%%%%%%%%
%%%%%%%%%%%%%%%%%%%%%%%%%%%%%%%%
\section*{Acknowledgments}

This work was supported by the Grant-in-Aid for Scientific Research on Scientific Research A (No.26247042 [KN]), 
Young Scientists B (No.\ 26800121 [KN]), Innovative Areas (No.\ 26104009 [KN], No.\ 15H05888 [KN]),
by World Premier International Research Center Initiative (WPI Initiative), MEXT, Japan,
by JSPS Research Fellowships for Young Scientists (Y.E. and K.M.),
and by the Program for Leading Graduate Schools, MEXT, Japan (Y.E.).

\appendix
%%%%%%%%%%%%%%%%%%%%%%%%%%%%%%%%
%%%%%%%%%%% Appendix %%%%%%%%%%%
%%%%%%%%%%%%%%%%%%%%%%%%%%%%%%%%

%%%%%%%%%%%%%%%%%%%%%%%%%%%%%%%%
\section{Mode Expansion}
\label{app:mode_exp}
%%%%%%%%%%%%%%%%%%%%%%%%%%%%%%%%

Here we summarize basic properties of the mode expansion of
the Higgs field:
\begin{align}
	h(x) = \int \frac{\dd^3 k}{\com{2 \pi a(t)}^{3/2}} 
	\com{
		\hat a_{\bm{k}} h_{\bm{k}} (t) e^{i \bm{k} \cdot \bm{x}} + \text{H.c.}
	},
	\label{eq:mode_exp_app}
\end{align}
where $\bm{k}$ is the comoving momentum and $a (t)$ is the scale factor.
Neglecting interaction terms,
we find the equation of motion for the wave function $h_{\bm k}$:
\begin{align}
	0 = \ddot h_{\bm {k}} (t) + \com{ \omega_{k; h}^2 (t) + \Delta (t)} h_{\bm k} (t),
	\label{eq:mode_eq_app}
\end{align}
where $\Delta \equiv - 9 H^2 /4 - 3 \dot H /2$, 
$H \equiv \dot a / a$ and 
$\omega_{k;h}(t)$ is the time dependent dispersion relation of Higgs.
We take the Wronskian of the Higgs field as
\begin{align}
	h_{\bm{k}} \dot h_{\bm{k}}^\ast - h_{\bm{k}}^\ast \dot h_{\bm{k}} = i,
\end{align}
which fixes the normalization of the wave function.
The wave equation leaves the Wronskian invariant since it is linear in $h_{\bm {k}}$.
Together with the canonical commutation relation of Higgs,
this normalization implies the following algebras for the creation/annihilation operator:
\begin{align}
	[\hat a_{\bm k}, \hat a^\dag_{\bm k'} ] = \delta ( \bm{k} - \bm{k'} ), 
	~~ [ \hat a_{\bm k}, \hat a_{\bm k'} ] = [ \hat a^\dag_{\bm k}, \hat a^\dag_{\bm k'} ] = 0.
\end{align}
Here note that there is redundancy of the expression in Eq~\eqref{eq:mode_exp_app}.
We can always rephrase Eq.~\eqref{eq:mode_exp_app} by another set of $(\hat{\tilde {\alpha}}_{\bm k}, \tilde h _{\bm k})$
satisfying 
$\tilde{\hat a}_{\bm k} = \alpha_{k} \hat a_{\bm k} + \beta_{k}^\ast \hat a^\dag_{- \bm k}$ and
$\tilde h_{\bm k} = \alpha_{k}^\ast h_{\bm k} - \beta_{k} h_{-\bm k}^\ast$
under $|\alpha_{k}|^2- |\beta_{k}|^2 = 1$.
This is the well-known Bogolyubov transform, ${\cal B}: (\hat a_{\bm k}, h_{\bm k}) \mapsto (\hat{\tilde {\alpha}}_{\bm k}, \tilde h _{\bm k})$, 
which leaves  the Wronskian, commutators and the norm $(h_{\bm k}, h_{- \bm k}^\ast) \cdot (\hat a_{\bm k}, \hat a_{- \bm k}^\dag)^t$ invariant.
By using this redundancy,
one can always take a basis which satisfies the following initial condition:\footnote{
 	The initial condition should be consistent with the normalization of Wronskian.
}
\begin{align}
	h_{\bm k} (t \to 0) \to \frac{1}{\sqrt{2 \omega_{k;h} (0)}},
	~~ \dot h_{\bm k} (t \to 0) \to - i \sqrt{ \frac{\omega_{k;h}(0)}{2}}.
\end{align}
Then, the initial vacuum state is annihilated by the corresponding operator $\hat a_{\bm k} \ket{0; \text{in}} = 0$.
We take this basis in the discussion given in this paper.
Here we have omitted contributions from the cosmic expansion $\sim {\cal O} (H^2 / \omega_{k;h}^2)$,
that is, we keep the leading order WKB result with respect to the cosmic expansion.

%%%%%%%%%%%%%%%%%%%%%%%%%%%%%%%%
\section{Renormalization in Classical Lattice Simulation}
\label{app:ren}
%%%%%%%%%%%%%%%%%%%%%%%%%%%%%%%%
In this appendix, we explain the renormalization procedure 
we have used in our classical lattice simulations.
We follow the procedure given in Refs.~\cite{Rajantie:2000nj,Berges:2013lsa}.

In the classical lattice simulations, we introduce initial gaussian fluctuations 
originating from the quantum fluctuations for each field.
Thus, effective mass terms are induced by these fluctuations.
To be concrete, we consider the Lagrangian~\eqref{eq:lag_ann} here.
Then, the initial mass terms of inflaton, Higgs and $\chi$ are given by
\begin{align}
	m_{\text{eff};\phi}^2(0)
	&=  m_\phi^2 + \frac{c^2}{2}\int_{{\bm k}}^{\Lambda} G_{F;h}(0,0;{\bm k}) 
	+ \delta m^2_{\Lambda;\phi}(0), \\
	m_{\text{eff};h}^2(0)
	&=  c^2\Phi_\text{ini}^2
	+ \frac{1}{2}\int_{{\bm k}} ^{\Lambda}
	\left[ c^2G_{F;\phi}(0,0; {\bm k}) 
	+ 3\lambda G_{F;h}(0,0; {\bm k})
	+ g_{h\chi}^2G_{F;\chi}(0,0; {\bm k})\right]
	+ \delta m^2_{\Lambda;h}(0), \\
	m_{\text{eff};\chi}^2(0)
	&= \frac{1}{2}\int_{{\bm k}}^{\Lambda}
	\left[ g_{h\chi}^2 G_{F;h}(0,0; {\bm k})
	+ 3g_{\chi\chi}^2 G_{F;\chi}(0,0; {\bm k})\right]
	+ \delta m^2_{\Lambda;\chi}(0),
\end{align}
where we have included the mass counter terms for inflaton, Higgs and $\chi$. 
These effective mass terms are UV sensitive since the statistical functions are initially given by
\begin{align}
	\frac{1}{2}G_{F;i}(0,0;{\bm k}) = \frac{1}{2\omega_{k;i}},
\end{align}
where $i = \phi, h$ and $\chi$. In the classical lattice simulations, 
the cut-off scale $\Lambda$ (or the regularization procedure)
is provided by the lattice discretization.
Thus, what we have to do here is to renormalize the cut-off scale dependence
by the mass counter terms.
We choose the mass counter terms such that the contributions from
the initial gaussian fluctuations are canceled by them, \textit{i.e.},
\begin{align}
	\delta m^2_{\Lambda;\phi}(0)
	&= -\frac{c^2}{2}\int_{{\bm k}}^{\Lambda} G_{F;h}(0,0;{\bm k}),
	\\
	\delta m^2_{\Lambda;h}(0)
	&= -\frac{1}{2}\int_{{\bm k}} ^{\Lambda}
	\left[ c^2G_{F;\phi}(0,0; {\bm k}) 
	+ 3\lambda G_{F;h}(0,0; {\bm k})
	+ g_{h\chi}^2G_{F;\chi}(0,0; {\bm k})\right],
	\\
	\delta m^2_{\Lambda;\chi}(0)
	&= -\frac{1}{2}\int_{{\bm k}}^{\Lambda}
	\left[ g_{h\chi}^2 G_{F;h}(0,0; {\bm k})
	+ 3g_{\chi\chi}^2 G_{F;\chi}(0,0; {\bm k})\right],
\end{align}
for the quartic coupling case.
We take the time evolution of the counter terms as 
\begin{align}
	\delta m^2_{\Lambda;i}(t) = \frac{m^2_{\Lambda;i}(0)}{a^{2}(t)},
\end{align}
where $i = \phi, h$ and $\chi$. 
This is because the physical cut-off scale $\Lambda_\text{cut}(t)$ 
evolves as $\Lambda_\text{cut}(t) = \Lambda/a(t)$ due to the cosmic expansion.
We have also renormalized the mass terms 
originating from the Higgs self coupling, Higgs-$\chi$ coupling and $\chi$ self coupling
in the same way for the curvature coupling case.

The renormalization procedure described here is unimportant for the analysis 
in Sec.~\ref{sec:higgs_inflaton}.
This is because the Higgs-inflaton couplings $c$ and $\xi$
are small in the case of our interest.
However, it is crucial for the study of the annihilation process given in Sec.~\ref{sec:annihilation}, 
where the couplings $g_{h\chi}$ and $g_{\chi\chi}$ are relatively large, of order $\mathcal{O}(0.1\mathchar`-1)$.
See Ref.~\cite{Berges:2013lsa} for more details on the importance of the renormalization procedure 
in the classical lattice simulations.

%%%%%%%%%%%%%%%%%%%%%%%%%%%%%%%%
\section{Thermalization after Inflation}
\label{app:thrm_aft_inf}
%!TEX root = ./inst_higgs_rh.tex

%Before going into details,
In this appendix, we summarize basic properties of thermalization after inflation
in the case of the reheating via a Planck-suppressed decay of inflaton.
Here we assume that the inflaton decays perturbatively,
and neglect the resonant production.
The thermalization process in this case is investigated in Refs.~\cite{Harigaya:2013vwa,Mukaida:2015ria}.\footnote{
	See also Ref.~\cite{Ellis:2015jpg}.
}
We follow the discussion given there.

Suppose that inflaton reheats the Universe via a Planck-suppressed decay,
$\Gamma_\phi = \tilde \Gamma_\phi m_\phi^3 / \Mpl^2$
with $\tilde \Gamma_\phi \ll 1$. 
In this case, the number density of radiation right after the decay of inflaton may be given by 
$n_h \sim \Gamma_\phi n_\phi / H \sim \tilde \Gamma_\phi m_\phi^3 (m_\phi t)^{-1}$, 
which is always smaller than the thermal one; so-called {\it ``under-occupied''} primaries~\cite{Kurkela:2011ti}.
The bottleneck process is in-medium collinear splittings of hard primaries~\cite{Kurkela:2011ti}
with the momentum of $p \sim m_\phi$. It is shown that, 
for $m_\phi t \lesssim \tilde t_\text{max} \equiv \alpha^{-16/5} \tilde \Gamma_\phi^{-3/5}$, %\lesssim {\cal O} (10^{3})$,
these hard primaries cannot participate in thermal plasma, and remain intact.
They may yield the following finite density corrections to the Higgs mass:
\begin{align}
	\left. \delta m^2_\text{th;h} \right|_\text{hard}
	\sim g^2 \tilde \Gamma_\phi m_\phi^2 \prn{ m_\phi t }^{-1}~~~
	\text{for}~~~ g_\text{eff} \abs{ h } \ll m_\phi.
	\label{eq:m_th_hard}
\end{align}
Here $g \sim y_t$ denotes  the electroweak gauge coupling and the top Yukawa  collectively,
to avoid unnecessary complications.
Note that $m_\phi t \gg 1$ is required since the inflaton should oscillate once at least so as to decay.

Though the hard primaries dominate the energy and number densities,
the soft population is produced via collinear splittings by them.
A quantum destructive interference effect prevents emission faster than
a time that it takes to resolve the overlaps between the parent and daughter,
that is $t \gtrsim k / k_\perp^2 \sim 1 / k \theta^2$.
In the medium, the daughter acquires transverse momentum by random collisions,
$k_\perp^2 \sim \hat q_\text{el} t$ with $\hat q_\text{el} \sim \alpha^2 n_h$ being the diffusion constant at that time.
Thus, for a given time $t$, there is an upper bound on the momentum, $ k \lesssim k_\text{form} \equiv \hat q_\text{el} t^2$.
While a medium induced cascade takes place below $k_\text{form}$ with a typical angle $\theta \lesssim  \alpha^{1/2}$,
a vacuum cascade may become relevant above $k_\text{form}$ with a minimum angle $\theta \gtrsim 1 / (kt)^{1/2}$~\cite{Kurkela:2014tla}.
On the one hand, if the formation momentum is lower than the Hubble parameter, $k_\text{form} \lesssim H$,
the finite density corrections to the Higgs mass may be dominated by the vacuum cascade spectrum.
On the other hand, for $k_\text{form} \gtrsim H$, the LPM-suppressed spectrum and the thermal-like spectrum
below $k_\text{max} \sim \alpha \tilde\Gamma_\phi^{1/2} m_\phi$ provide dominant corrections to the Higgs mass.
After a characteristic time scale, $m_\phi t > \tilde t_\text{soft} \equiv \alpha^{-3} \tilde \Gamma_\phi^{- 1/2}$,
the soft populations are thermalized among themselves.
Eventually, for $m_\phi t > \tilde t_\text{max} \equiv \alpha^{-16/5} \tilde \Gamma_\phi^{-3/5}$,
the radiation, including hard primaries, is thermalized against the expansion of the Universe,
and follows the standard evolution.
Thus, the soft population may yield the following corrections to the Higgs mass term~\cite{Harigaya:2013vwa,Mukaida:2015ria}:\footnote{
	Owing to the Fermi-Dirac statistics, the soft sector is dominated by bosons, {\it i.e.}~SM gauge bosons.
}
\begin{align}
	\left. \delta m_\text{th;h}^2 \right|_\text{soft} 
	\sim g^2
	\begin{cases}
		\alpha \tilde \Gamma_\phi m_\phi^2 &\text{for}~~g \abs{h} \ll k_\text{max}\sim H;~~ m_\phi t < \tilde t_\text{el}, \\[1em]
		\alpha \tilde \Gamma_\phi m_\phi^2 \prn{\cfrac{\tilde t_\text{el}}{m_\phi t}}^\frac{1}{2}
		&\text{for}~~g \abs{h} \ll k_\text{max} \sim \alpha \tilde \Gamma_\phi^\frac{1}{2} m_\phi;~~ \tilde t_\text{el} < m_\phi t <  \tilde t_\text{soft}, \\[1.5em]
		\alpha^2 \tilde \Gamma_\phi m_\phi^2 \prn{ \cfrac{m_\phi t}{\tilde t_\text{soft} }}^2
		&\text{for}~~g \abs{h} \ll k_\text{max} \sim \alpha^4 \tilde \Gamma_\phi m_\phi (m_\phi t);~~ \tilde t_\text{soft} < m_\phi t <  \tilde t_\text{max}, \\[1.5em]
		\alpha^\frac{8}{5}  \tilde \Gamma_\phi^\frac{4}{5} m_\phi^2 \prn{ \cfrac{\tilde t_\text{max}}{m_\phi t} }^\frac{1}{2}
		&\text{for}~~g \abs{h} \ll k_\text{max}\sim \tilde \Gamma_\phi^\frac{1}{4} m_\phi (m_\phi t)^{-\frac{1}{4}};~~ \tilde t_\text{max} < m_\phi t <  \tilde t_\text{RH}.
	\end{cases}
	\label{eq:m_th_soft}
\end{align}
Here the time after which medium effects dominate splittings is defined as 
$\tilde t_\text{el}= \alpha^{-1} \tilde \Gamma_\phi^{- 1/2}$;
the time after which the soft populations are thermalized is
$\tilde t_\text{soft}= \alpha^{-3} \tilde \Gamma_\phi^{- 1/2}$;
the time after which the radiation is thermalized is
$\tilde t_\text{max} = \alpha^{-16/5} \tilde \Gamma_\phi^{-3/5}$;
and the time when the reheating is completed is 
$\tilde t_\text{RH} = \tilde \Gamma_\phi^{-1} \Mpl^2 / m_\phi^2$.
%%%%%%%%%%%%%%%%%%%%%%%%%%%%%%%%

%%%%%%%%%%%%%%%%%%%%%%%%%%%%%%%%
%%%%%%%%%%% References %%%%%%%%%%%
%%%%%%%%%%%%%%%%%%%%%%%%%%%%%%%%
\small
\bibliography{ref}

\end{document}